\documentclass[a4paper,fleqn]{cas-sc}

\usepackage[numbers]{natbib}
\usepackage[utf8]{inputenc}
\usepackage{amssymb}
\usepackage{hyperref}
\usepackage{array}
\usepackage{caption}
\usepackage{graphicx}
\usepackage{subcaption}
\usepackage[dvipsnames]{xcolor}

\def\tsc#1{\csdef{#1}{\textsc{\lowercase{#1}}\xspace}}
\tsc{WGM}
\tsc{QE}

\begin{document}
\let\WriteBookmarks\relax
\def\floatpagepagefraction{1}
\def\textpagefraction{.001}

\shorttitle{Refining ADHD Diagnosis with EEG: The Impact of Preprocessing and Temporal Segmentation on Classification Accuracy}    

\shortauthors{S. García-Ponsoda, A. Maté, J. Trujillo}  

\title [mode = title]{Refining ADHD Diagnosis with EEG: The Impact of Preprocessing and Temporal Segmentation on Classification Accuracy}  



%

\author[inst1,inst2]{Sandra García-Ponsoda}[orcid=0000-0002-0739-6680]

\cormark[1]

\fnmark[1]

\ead{sandra.gp@ua.es}

\affiliation[inst1]{organization={Lucentia Research Group - Department of Software and Computing Systems},
            university={University of Alicante},
            addressline={Rd. San Vicente s/n}, 
            city={San Vicente del Raspeig},
            postcode={03690}, 
            country={Spain}}
\affiliation[inst2]{organization={ValgrAI - Valencian Graduate School and Research Network of Artificial Intelligence},
            addressline={Camí de Vera s/n}, 
            city={Valencia},
            postcode={46022},
            country={Spain}}

\author[inst1]{Alejandro Maté}[orcid=0000-0001-7770-3693]

\ead{amate@dlsi.ua.es}

\author[inst1,inst2]{Juan Trujillo}[orcid=0000-0003-0139-6724]

\ead{jtrujillo@dlsi.ua.es}

\cortext[1]{Corresponding author}

\fntext[1]{Postal address: University Institute for Computing Research II, University of Alicante, Rd. San Vicente s/n, San Vicente del Raspeig, 03690, Spain}


\begin{abstract}
Background:
EEG signals are commonly used in ADHD diagnosis, but they are often affected by noise and artifacts. Effective preprocessing and segmentation methods can significantly enhance the accuracy and reliability  of ADHD classification.

Methods:
We applied filtering, ASR, and ICA preprocessing techniques to EEG data from children with ADHD and neurotypical controls. The EEG recordings were segmented, and features were extracted and selected based on statistical significance. Classification was performed using various EEG segments and channels with Machine Learning models (SVM, KNN, and XGBoost) to identify the most effective combinations for accurate ADHD diagnosis.

Results:
Our findings show that models trained on later EEG segments achieved significantly higher accuracy, indicating the potential role of cognitive fatigue in distinguishing ADHD. The highest classification accuracy (86.1\%) was achieved using data from the P3, P4, and C3 channels, with key features such as Kurtosis, Katz fractal dimension, and power spectrums in the Delta, Theta, and Alpha bands contributing to the results.

Conclusions:
This study highlights the importance of preprocessing and segmentation in improving the reliability of ADHD diagnosis through EEG. The results suggest that further research on cognitive fatigue and segmentation could enhance diagnostic accuracy in ADHD patients.
\end{abstract}

\begin{graphicalabstract}
\includegraphics[width=\textwidth]{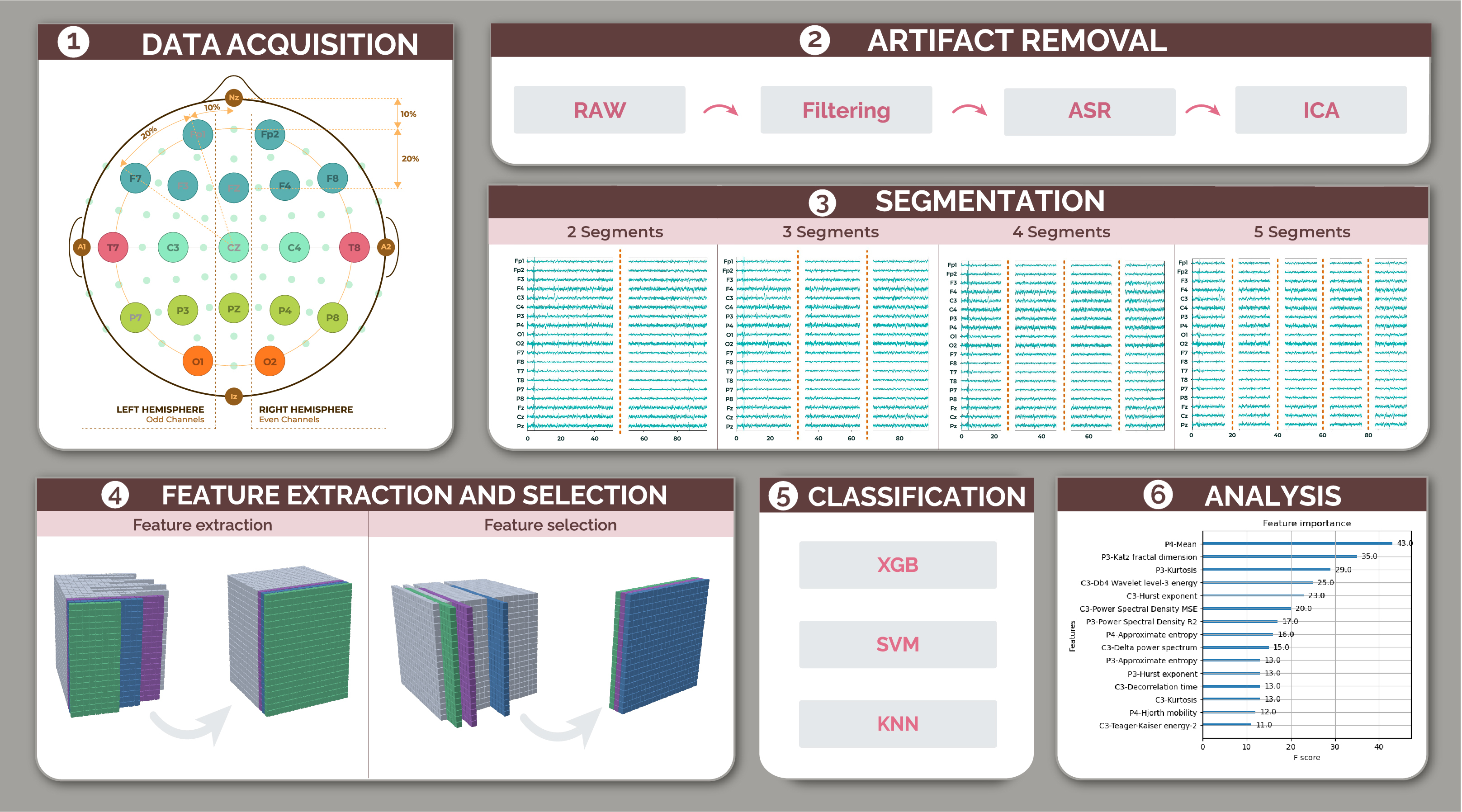}
\end{graphicalabstract}

\begin{highlights}
    \item Comprehensive study showing the necessity of effective preprocessing in EEG-based ADHD classification.
    \item High classification accuracy (86.1\%) using data from only three EEG channels (P3, P4, and C3)
    \item Segmenting EEG recordings reveals increased accuracy in later segments, likely due to ADHD-related cognitive fatigue.
    \item Key features identified include Kurtosis, Katz fractal dimension, and power density in Delta, Theta, and Alpha bands.
\end{highlights}


\maketitle

\section{Introduction}\label{sec:Intro}

Attention-Deficit/Hyperactivity Disorder (ADHD) is a chronic neurodevelopmental disorder that affects approximately 7.6\% of children and 5.6\% of adolescents globally \cite{salari2023global}, as well as 6.76\% of adults \cite{song2021prevalence}. Despite extensive research, the exact cause of ADHD remains unidentified, as noted by the American Psychiatric Association \cite{diagnostic1994statistical}.

ADHD is a chronic disorder that significantly impacts academic performance, relationships, family well-being, and mental health \cite{harpin2005effect}. It affects attention, impulse control, activity levels, and executive functions. Individuals with ADHD frequently move their limbs, have difficulty sitting still, struggle to engage quietly in recreational activities, talk excessively, and/or display impatience \cite{alma9918660570001341}. Untreated ADHD can lead to serious problems such as low self-esteem, school dropout, criminal behavior, and addiction \cite{lola2019attention, el2023adult}. Therefore, early, accurate, and reliable diagnosis is crucial.

The diagnosis of ADHD traditionally involves expert interviews, Continuous Performance Test (CPT) results, and the Diagnostic and Statistical Manual of Mental Disorders, Fifth Edition (DSM-V) guidelines \cite{alma9918660570001341}. However, these methods often lack sufficient sensitivity and specificity \cite{marshall2021diagnosing}. Advanced techniques are now being proposed to improve diagnostic accuracy and provide deeper insights into ADHD.

Studies have employed neuroimaging techniques such as Magnetic Resonance Imaging (MRI) \cite{karavallil2023alterations, lohani2023adhd}, Magnetoencephalography (MEG) \cite{serrallach2022neuromorphological}, and Positron Emission Tomography (PET) \cite{millevert2023resting} to diagnose ADHD. However, these methods are expensive, bulky, data-intensive, and sensitive to subject movement, as even minor movements can lead to unusable outputs, necessitating the repetition of neuroimaging procedures.

There are also other techniques that gather physiological signals, such as Electroencephalograms (EEG), which capture the electric fields generated by current sources in the brain. EEG is one of the most commonly used methodologies in ADHD research \cite{hassan2024convolutional, latifi2024siamese, jahani2024efficient} and in many other fields \cite{degirmenci2024eeg, zhang2024review, altaheri2023deep}, primarily because it is inexpensive, convenient, and has been commercially developed into portable and wireless designs \cite{wolpaw2012brain}. Additionally, unlike neuroimaging, EEG is relatively tolerant to movement, meaning that a recording remains valid even if the subject moves briefly during the session. EEG is recorded using sensors, also known as channels, placed on a helmet following standard arrangements, such as the 10-20, 10-10, or 10-5 systems \cite{jurcak200710}. However, the signals typically contain noise and external information (such as muscle movements, blinks, and electrical grid interference, among others), which often interfere with the detection and measurement of signals relevant to the study's objective. Therefore, it is crucial to detect and remove these unwanted signals before analysis, feature extraction, or modelling to ensure accurate and meaningful results, as demonstrated in most EEG-based studies \cite{bigdely2015prep, nolan2010faster, mumtaz2021review}.

EEG signal preprocessing is critical in extracting meaningful brainwave patterns, removing noise and artifacts that may obscure important information. The choice of preprocessing techniques directly affects the signal-to-noise ratio, impacting the performance of ML classifiers. This study investigates how different preprocessing pipelines affect the accuracy of ADHD classification using a publicly available EEG dataset.

Currently, researchers are focusing on differentiating individuals with ADHD from those with neurotypical development (TD) using artificial intelligence \cite{alsharif2024diagnosis, cura2024detection, sharma2023classification}. However, for the differences between both groups (ADHD and TD children) to be reliable and appropriate, experts in the field must understand them. Explainability in Machine Learning (ML) models is particularly important in healthcare settings, where understanding model decisions can guide clinical diagnostics. This study not only explores preprocessing techniques but also identifies which EEG features and channels most contribute to classification accuracy. This transparency helps ensure that the developed models can be trusted and interpreted by clinicians for ADHD diagnosis.

Recent studies have focused on classifying ADHD using EEG signals. However, according to \cite{loh2022automated}, only one public database with a significant number of subjects is available for research. Several recent works have utilized this database. The lack of an identified cause for ADHD has led to each of these studies applying different preprocessing methods. Table \ref{tab:rel_works} presents a summary of the preprocessing methods used in the most recent studies that have utilized the public database \cite{nasrabadi2020eeg}, which is also employed in this study.

\begin{table}[h]
    \centering
    \begin{tabular}{ c >{\centering\arraybackslash}m{1.5cm} c >{\centering\arraybackslash}m{2cm} c c }
        \toprule
        \textbf{Study} & \textbf{Visual Inspection} & \textbf{Filtering} & \textbf{Line Noise Removal} & \textbf{Artifact Removal} & \textbf{Windowing} \\
        \midrule
        Abedinzadeh et. al\cite{abedinzadeh2023potential} &  No & [1, 48] Hz & - & Clean\_rawdata + ICA & 2s \\
        
        Lin et. al \cite{lin2023measurement} &  No & [1, 48] Hz & - & ICA & - \\
        
        Gu et. al\cite{gu2023detection} &  No & [0.5, 45] Hz & CleanLine & ICA & - \\
        
        Bakhtyari et. al\cite{bakhtyari2022adhd} &  No & [4, 40] Hz & - & ICA & - \\
        
        Sanchis et. al\cite{sanchis2024novel} &  No & [0.5, 60] Hz & Notch filter (50Hz) & ASR (clean\_rawdata) & 2s \\
        
        Atila et. al\cite{atila2023lsgp} &  No & [0, 60] Hz & Notch filter (50Hz) & - & - \\
        
        Abbas et. al\cite{abbas2021effective} &  No & [0.5, 45] Hz & - & - & 3s \\
        
        Khare et. al\cite{khare2022vhers} &  No & [0, 60] Hz & Notch filter (50Hz) & - & 4s \\
        
        Khare et. al\cite{khare2023explainable} & No & [0.1, 60] Hz & Notch filter (50Hz) & - & 4s \\
        
        Maniruzzaman et. al\cite{maniruzzaman2023optimal} & No & - & - & - & - \\
        
        Loh et. al\cite{loh2023adhd} & No & - & - & - & - \\
        
        Kasim \cite{kasim2023identification} & No & - & - & - & 10s \\
        
        Ge et. al\cite{ge2023symbolic} &  No & - & - & - & length=1000 \\
        
        Barua et. al\cite{barua2022tmp19} &  No & - & - & - & length=5 \\
        
        Ekhlasi et. al\cite{ekhlasi2021direction} & Yes & [0.5,48] Hz & CleanLine & ICA & 8s \\
        
        Talebi et. al\cite{talebi2022investigating} & Yes & [0.5, $+\infty$) Hz & CleanLine & ICA & - \\
        
        Chauhan et. al\cite{chauhan2023regional} & Yes & [0.5, 48] Hz & CleanLine & ICA & 8s \\
        \bottomrule
    \end{tabular}
    \caption{Summary of the preprocessing methods used in related studies that utilize the same database as our study.}
    \label{tab:rel_works}
\end{table}

From Table \ref{tab:rel_works}, we can identify four important steps in EEG signal preprocessing: filtering, handling line noise, and artifact removal. Additionally, we have included a column for visual inspection, as we believe it is essential to automate preprocessing methods; otherwise, they require significant effort and are time-consuming. We also included a column related to dividing EEG recordings into windows, a widespread practice. The filtering column indicates the frequency range used in each study, while the line noise column indicates whether a specific filter was used to remove electrical grid interference. The artifact removal column specifies the algorithms or methods applied to separate artifacts from brain signals.

\cite{ekhlasi2021direction, talebi2022investigating, chauhan2023regional} are the only studies that perform an initial visual inspection to eliminate large artifacts, such as muscle or eye movements. On the other hand, studies like \cite{maniruzzaman2023optimal, loh2023adhd, kasim2023identification, ge2023symbolic, barua2022tmp19} do not apply any preprocessing to conduct their experiments. \cite{atila2023lsgp, khare2022vhers, khare2023explainable} rely solely on filters to capture frequencies up to 60 Hz and apply a 50 Hz notch filter to remove electrical noise. \cite{bakhtyari2022adhd, gu2023detection, lin2023measurement} use filters to eliminate large artifacts, followed by Independent Component Analysis (ICA) to remove remaining artifacts. Instead of ICA, \cite{sanchis2024novel} employs the Artifact Subspace Reconstruction (ASR) algorithm, available in EEGLAB's clean\_rawdata() functionality. Finally, \cite{abedinzadeh2023potential} is the only study that performs all steps: applying a 1–40 Hz band-pass Finite Impulse Response (FIR) filter to remove line noise and low frequencies, using the clean\_rawdata() plug-in to automatically eliminate visible artifacts from electrode displacement, and ICA to eliminate remaining unwanted artifacts such as muscle movements, heartbeats, or eye blinks.

The main objective of our study is to improve the reliability of ADHD detection through preprocessing of EEG signals. To achieve this, the questions guiding our work are the following:
\begin{itemize}
    \item How does the choice of EEG signal preprocessing influence the classification of individuals with ADHD? 
    \item Is that choice decisive for the results?
    \item Does it make a significant difference?
\end{itemize}

Following preprocessing, we investigate whether specific segments of EEG recordings provide more distinguishing features between ADHD and TD children. By segmenting the data, we aim to identify critical time windows that could improve classification performance, demonstrating that effective preprocessing and time-based analysis are key to more reliable ADHD diagnosis. The question guiding this aim is:
\begin{itemize}
    \item Could segmenting the recording help identify important moments or patterns that improve classification?
\end{itemize}
Finally, as we have worked with explainable models, we identify the most relevant features and channels for classification, aiming to provide a transparent and interpretable model for clinicians.

We use a public database of EEG signals from children with ADHD to achieve our objectives. We compare three preprocessing pipelines with different data cleaning techniques and segment the EEG records to identify the most discriminative parts. Due to the exponential increase in the number of $k$-element combinations of 19 channels, without repetition, which is given by $C_{19,k} = \frac{19!}{k!(19-k)!}$, we limit our selection to k = 1, 2, and 3 channels. Regarding the time required for training the models, when training all model combinations for the 3-channel sets, the process took approximately 5 days. Therefore, increasing the number of channels by just one would result in an estimated training time of around 20 days.


This study makes several important contributions to the field of ADHD diagnosis using EEG data. First, it demonstrates the critical role of proper preprocessing (filtering, ASR, and ICA) in ensuring reliable and unbiased results, avoiding the artificially high accuracies that can result from unclean data. Additionally, the study introduces a novel segmentation approach, showing that later segments of EEG recordings provide significantly higher classification accuracy, offering new insights into attention fatigue in ADHD subjects. Furthermore, the paper identifies key EEG channels (P3 and P4) and critical features such as Kurtosis, Katz fractal dimension, and various power spectrums (Delta, Theta, Alpha), which are most relevant for ADHD classification. Remarkably, the study achieves high classification accuracy (up to 86\%) using only three channels, streamlining data collection and reducing computational complexity. Finally, by sharing the complete source code, the study emphasizes openness and reproducibility, facilitating further research in this area.

This paper is structured as follows: in Section \ref{sec:matandmet}, we describe the dataset (Section \ref{sec:dataset}), the preprocessing methods (Section \ref{sec:prep}), how the data are segmented (Section \ref{sec:chunks}), the extraction and selection of features (Sections \ref{sec:feat-extraction} and \ref{sec:feat-selection}), and the classification and training process (Section \ref{sec:clf-models}). In Section \ref{results}, we present and analyze the results. In Section \ref{disc}, we discuss these findings, and finally, in Section \ref{concl}, we conclude the paper.

\section{Materials and methods}\label{sec:matandmet}

\begin{figure}[ht]
    \centering
        \includegraphics[width=\textwidth]{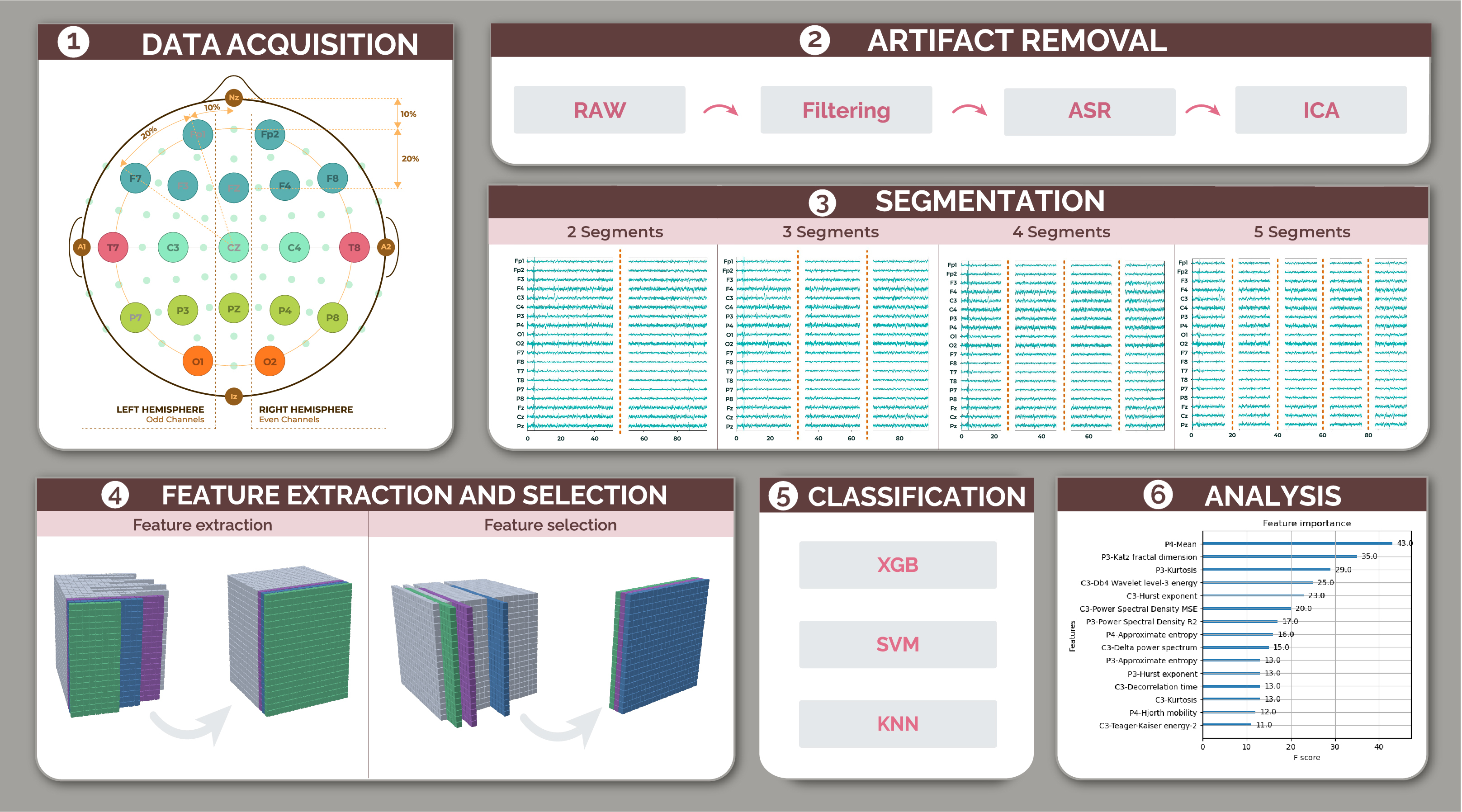}
    \caption{Pipeline illustrating how data have been acquired, preprocessed, classificated and analized.}
    \label{fig:pipeline}
\end{figure}

This section contains a description of the dataset used. Following that, we describe how the dataset is preprocessed, detailing all the techniques employed. We also explain how we measure the existing noise and segment the EEG recordings. Next, the extraction and selection of features are described. Finally, we present the classification models used and explain how they are trained. Figure \ref{fig:pipeline} illustrates the entire process.

\subsection{Dataset}\label{sec:dataset}

\begin{figure}[ht]
    \centering
        \includegraphics[width=0.5\textwidth]{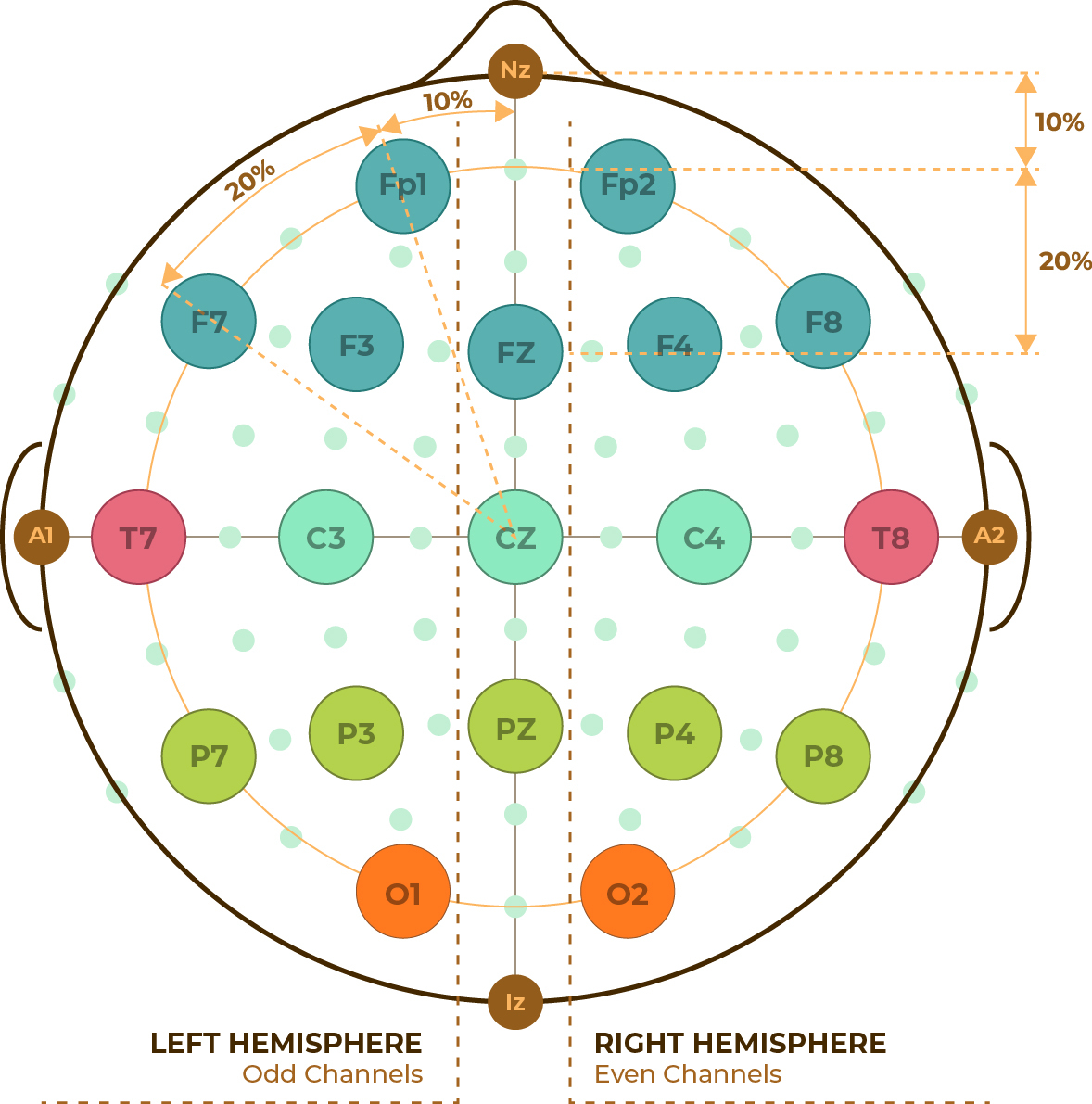}
    \caption{10-20 International electrode placement layout using 19 EEG channels. The color grouping indicates the different regions: frontal (turquoise), central (aquamarine), temporal (pink), parietal (green), and occipital (orange). Created by authors.}
    \label{fig:10-20-placement}
\end{figure}

In this study, we utilize one of the few public EEG datasets available for children with ADHD \cite{nasrabadi2020eeg}. The dataset comprises 61 children with ADHD (48 boys and 13 girls, mean age $9.62 \pm 1.75$) and 60 TD children (50 boys and 10 girls, mean age $9.85 \pm 1.77$) \cite{ekhlasi2021direction}, all right-handed. ADHD diagnoses were made by experienced psychiatrists according to DSM-IV guidelines. None of the TD children had a history of psychiatric disorders, epilepsy, or high-risk behaviors. EEG signals were recorded using a 19-channel helmet placed on the child’s scalp, labeled Fp1, Fp2, Fz, Cz, Pz, C3, T3, T4, F3, F4, F7, F8, P3, P4, T5, T6, O1, and O2, as shown in Figure \ref{fig:10-20-placement}.

\begin{figure}[ht]
    \centering
        \includegraphics[width=0.5\textwidth]{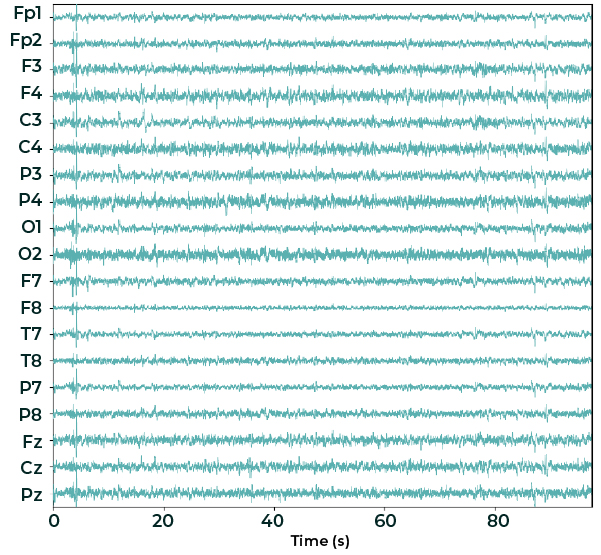}
    \caption{Example of EEG recording from a randomly selected subject.}
    \label{fig:eeg-example}
\end{figure}

The channel placement layout follows the international 10-20 system \cite{acharya2016american}. The recording helmet has a sampling frequency of 128 Hz, capturing 128 samples per second. Each experiment involves a child wearing the EEG helmet while seated in front of a screen displaying visual stimuli. An image appears with a random number of characters, ranging from 5 to 16, and the child is asked to enumerate the characters. After the child's response, another image is presented, continuing until 20 images have been shown. Thus, each experiment varies in length depending on the child’s performance. This experiment is performed once per subject, resulting in 121 EEG recordings of varying lengths. Figure \ref{fig:eeg-example} illustrates the structure of the EEG recordings of each child.

\subsection{Data preprocessing}\label{sec:prep}

In this section, we describe the steps taken to prepare the raw EEG signals prior to feature extraction and selection. As the goal of our study is to understand the impact of noise removal on classification outcomes, we employ three different preprocessing methods for comparison. These methods are based on the most commonly used techniques in the literature, as mentioned in Section \ref{sec:Intro}. Each method removes varying amounts of noise from the data. The following subsections explain how the noise was removed and what tools were used (Sections \ref{sec:filtering}, \ref{sec:ASR}, and \ref{sec:ICA}). Additionally, Section \ref{sec:noise-measurement} quantifies the noise removed by each preprocessing method.

We begin with 121 matrices, one for each subject. Each matrix has dimensions of $19 \times T_{i}$, where $T_{i}$ represents the duration of the experiment for subject $i$, with $i \in \{0, 120\}$. These matrices contain a row for each EEG channel and a column representing the numerical values collected by the channels at each time instant. Given the helmet's sampling frequency of 128 Hz, 128 samples are recorded per second. It is important to note that each subject's recording length varies.

\subsubsection{Filtering process}\label{sec:filtering}

Filtering is a fundamental preprocessing step in EEG signal analysis. The purpose of filtering is to attenuate or remove specific frequencies that are not of interest or are considered noise.

Since the specific cause of ADHD is not yet known, in this study, we aim to capture the widest possible frequency spectrum. According to Nyquist's theorem, with a sampling frequency of 128 Hz, we can reconstruct signals up to 64 Hz. Additionally, in regions like Europe, Asia, Africa, and Oceania, power-line interference occurs at 50 Hz. As shown in Figure \ref{fig:prep-raw-example}, this interference affects the 40-60 Hz range. Therefore, we apply a band-pass filter between 0.5 and 40 Hz to capture frequency bands such as delta (0.5-4 Hz), theta (4-8 Hz), alpha (8-13 Hz), beta (13-30 Hz), and gamma (>30 Hz). This filtering also removes large artifacts, such as blinking and muscle movements. The filter is implemented as a Finite Impulse Response (FIR) filter using MATLAB.

\subsubsection{Artifact Subspace Reconstruction process}\label{sec:ASR}

Artifact removal is another key step in EEG preprocessing, although, as discussed in Section \ref{sec:Intro}, only \cite{abedinzadeh2023potential} and \cite{sanchis2024novel} employ it. We chose the Artifact Subspace Reconstruction (ASR) method \cite{kothe2014artifact} for artifact removal due to its automatic capability to identify and utilize clean segments of EEG recordings to remove noisy components. Additionally, ASR preserves the EEG signal more effectively than aggressive filtering methods. The ASR process consists of five key steps. First, ASR selects clean portions of EEG signals based on the distribution of signal variance, aiming to define a clean reference subspace. Second, ASR decomposes the EEG signal into subspaces using Principal Component Analysis (PCA). Third, ASR establishes rejection criteria based on the signal variance distribution in the principal component space. Fourth, components containing artifacts are identified and selected or rejected based on the predefined criteria. Finally, the identified artifact components are reconstructed by projecting them back onto the clean reference subspace (created in the first step), preserving the underlying neural signals \cite{chang2018evaluation}. The technical and mathematical details of these steps are discussed comprehensively in \cite{kothe2014artifact} and \cite{mullen2015real}.

We utilized the open-source EEGLAB plugin \textit{clean\_rawdata()} to perform ASR \cite{delorme2004eeglab, makoto2021cleanrawdata}. Rather than removing segments containing artifacts, we reconstructed them. To create the second dataset, we applied the ASR algorithm to the filtered dataset, described in Section \ref{sec:filtering}.

\subsubsection{Independent Component Analysis process}\label{sec:ICA}

EEG measures the electrical potential difference between two electrodes, commonly referred to as channels. Each channel captures voltage fields generated by nearby brain sources, along with non-brain sources. Independent Component Analysis (ICA) is a computational technique used to separate these brain sources. ICA achieves this by decoupling the problem of source identification from that of source localization \cite{makeig1995independent}. As described in \cite{wolpaw2012brain}, ICA functions as a spatial filter and can be represented as a linear combination of channels in matrix notation: $Y = WX$ or in a matrix form as:

\begin{align}
    \begin{bmatrix}
        y_{11} & y_{12} & \cdots  & y_{1P}\\
        y_{21} & y_{22} &         &       \\
        \vdots &        & \ddots  &       \\
        y_{M1} &        &         & y_{MP}
    \end{bmatrix}
    =
    \begin{bmatrix}
        w_{11} & w_{12} & \cdots  & w_{1N}\\
        w_{21} & w_{22} &         &       \\
        \vdots &        & \ddots  &       \\
        w_{M1} &        &         & w_{MN}
    \end{bmatrix}
    \begin{bmatrix}
        x_{11} & x_{12} & \cdots  & x_{1P}\\
        x_{21} & x_{22} &         &       \\
        \vdots &        & \ddots  &       \\
        x_{N1} &        &         & x_{NP}
    \end{bmatrix}
\end{align}

Where the matrix $X$ consists of $P$ digital signal samples from $N$ channels, the matrix $W$ contains a set of $N$ channel weights that form a particular spatial filter, and the matrix $Y$ represents the spatially filtered channels ($M$ spatially filtered channels $\times$ $P$ samples). ICA seeks to determine the weight matrix $W$ that produces independent channels in $Y$, meaning that the joint probability distribution of channels $x$ ans $y$ $F_{XY}(x,y) = Pr\{ X \leq x, Y \leq y \}$ must satisfy that $F_{XY} (x,y) = F_{X}(x)F_{Y}(y)$.

It is important to note that the maximum number of independent sources ICA can identify is equal to the number of channels in $X$, i.e., $N$. As suggested by \cite{wolpaw2012brain}, if the number of original sources exceeds the number of channels, it is common to use additional cleaning techniques prior to ICA to remove irrelevant sources. In our case, we applied a FIR filter (Section \ref{sec:filtering}) and the ASR algorithm (Section \ref{sec:ASR}) to remove noticeable artifacts before using the ICA method. Once the independent components are identified by ICA, we use the ICALabel package to classify the components based on their source: brain, muscle, ocular, heartbeats, line noise, channel noise, and others \cite{pion2019iclabel}. We retain only the independent components labeled as ``brain components" and reconstruct the EEG signal from these components. In this manner, we create the third cleaned dataset for comparison purposes.

\subsubsection{Noise measurement}\label{sec:noise-measurement}

\begin{figure}[ht]
\centering
\begin{subfigure}{0.35\textwidth}
    \centering
        \includegraphics[width=0.9\textwidth]{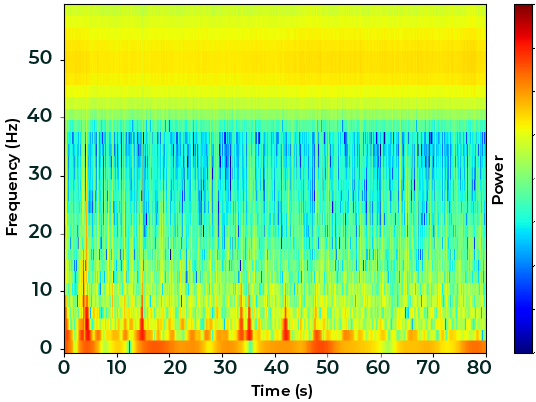}
    \caption{Time-frequency plot of \textit{Raw} data.}
    \label{fig:prep-raw-example}
\end{subfigure}
\begin{subfigure}{0.35\textwidth}
    \centering
        \includegraphics[width=0.9\textwidth]{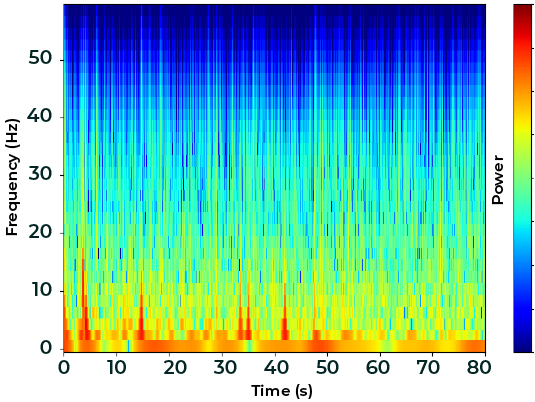}
    \caption{Time-frequency plot of \textit{Filtered} data.}
    \label{fig:prep-filt-example}
\end{subfigure}
\begin{subfigure}{0.35\textwidth}
    \centering
        \includegraphics[width=0.9\textwidth]{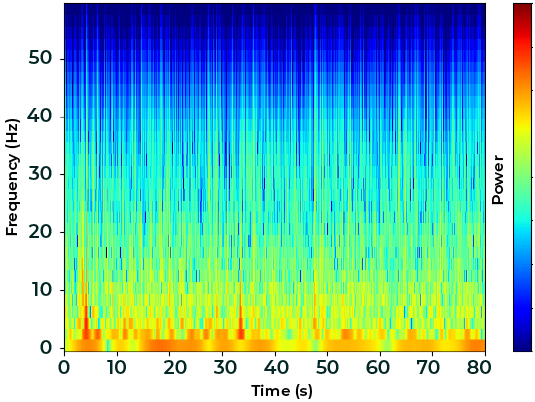}
    \caption{Time-frequency plot of \textit{ASR} data.}
    \label{fig:prep-asr-example}
\end{subfigure}
\begin{subfigure}{0.35\textwidth}
    \centering
        \includegraphics[width=0.9\textwidth]{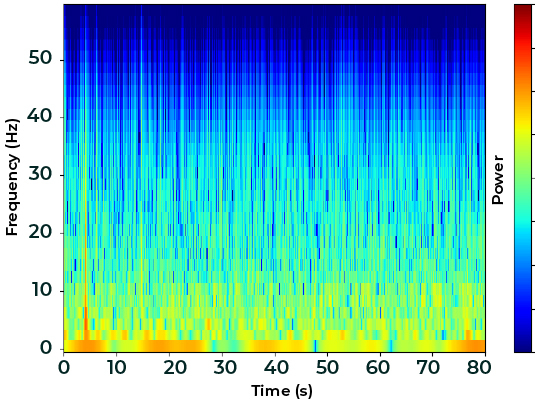}
    \caption{Time-frequency plot of \textit{ICA} data.}
    \label{fig:prep-ica-example}
\end{subfigure}    
\caption{Time-frequency plots of a randomly selected subject's experiment. Each plot corresponds to a different cleaning process.}
\label{fig:example-preprocessings}
\end{figure}

It is possible to graphically observe that each of the chosen cleaning methods removes a different amount of noise. Figure \ref{fig:example-preprocessings} shows the time-frequency plots of the Fp1 channel from a randomly selected subject. In Figure \ref{fig:prep-raw-example}, we present the raw data, where power-line noise is evident around the $(40,60)$ Hz range, along with artifacts at the bottom. Figure \ref{fig:prep-filt-example} displays the data after filtering, where the power-line noise has been removed, revealing more artifacts. Figure \ref{fig:prep-asr-example} illustrates the data after applying ASR, showing attenuation of several artifacts. Finally, Figure \ref{fig:prep-ica-example} shows the data post-ICA, where most artifacts are nearly eliminated.

Following this trend, we hypothesize that the ICA method removes more noise than the ASR algorithm, and the ASR algorithm eliminates more noise than the filtering technique. However, it is important to quantify the amount of noise removed to confirm this behavior across all subjects and channels in the dataset. We use the Signal-to-Noise Ratio (SNR) to quantify the removed noise, as it is a widely used measure across various fields \cite{klug2024optimizing, reddy2024nonlinear, zhang2024adaptive}. SNR is defined as the ratio between the signal power and the noise power corrupting it. In the context of EEG, SNR quantifies how much of the recorded signal represents actual brain activity versus noise from external and internal sources such as muscle movements, electrical interference, blinking, or heartbeats. SNR is calculated as the ratio of signal power to noise power:

\begin{equation}
    \text{SNR} = \frac{\text{PWR}_{s}}{\text{PWR}_{n}}
\end{equation}

Where $\text{PWR}_{s}$ represents the power of the remaining EEG signal, and $\text{PWR}_{n}$ is the power of the identified noise. We calculated the SNR for each of the three cleaning methods (filtering, ASR, and ICA). The raw EEG data served as a reference for determining the amount of noise removed by each method. For this calculation, the raw signal power was used as $PWR_{s}$, and the difference between the raw signal power and the preprocessed signal power was used as $PWR_{n}$.

In terms of interpretability, a higher SNR indicates that the amount of noise removed is minimal compared to the remaining signal. In contrast, a lower SNR suggests that the noise removed is substantial relative to the remaining signal.

\begin{figure}[ht]
    \centering
    \includegraphics[width=0.8\textwidth]{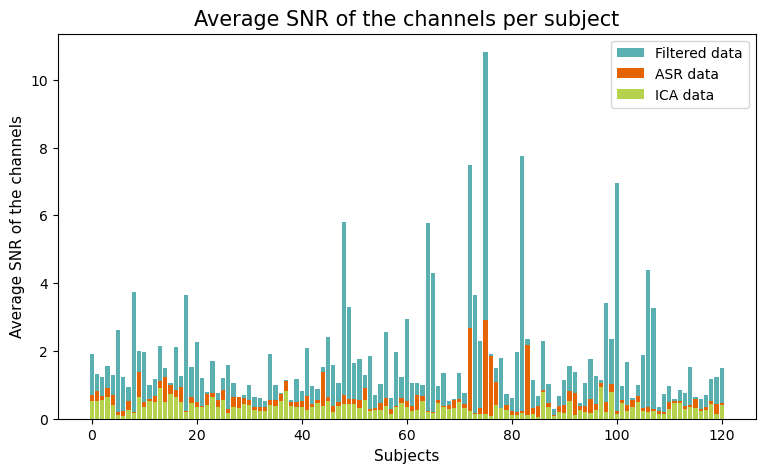}
    \caption{Average SNR of the 19 channels per subject according to the cleaning method used (Filtering, ASR, and ICA).}
    \label{fig:snr}
\end{figure}

Figure \ref{fig:snr} displays the average SNR of the channels for each subject in the dataset. It is evident that the ICA technique is the most effective at eliminating noise, followed by the ASR algorithm, while the filtering technique removes the least amount of noise.

\subsection{Data segmentation}\label{sec:chunks}

In Table \ref{tab:rel_works}, we observe that the only segmentation technique employed is windowing. Windowing involves temporally dividing the EEG signals into windows of $n$ seconds, where $n$ belongs to the set of natural numbers ($n \in \mathbb{N}$). However, in this section, we propose a novel step in the EEG preprocessing pipeline for ADHD. We decided to split the EEG recordings to search for segments where subjects exhibit the most differences between the two groups. This approach shifts away from treating all segments uniformly and emphasizes the significance of specific temporal points from which each segment is derived.

To achieve this, we divided the EEG recordings into 2, 3, 4, 5, and 20 equal segments. Additionally, we retained the entire recordings for comparison. We chose to split the data into 2, 3, 4, and 5 segments to avoid making the segments too short, as the average length of the recordings is approximately $49.33$ seconds. The decision to divide the recordings into 20 segments is based on the number of images shown in all experiments, as indicated in \cite{ekhlasi2021direction}. Since the dataset does not provide timestamps for image presentations, we divided the EEG recordings into 20 equal parts.

\begin{figure}[ht]
\centering
\begin{subfigure}{0.4\textwidth}
    \centering
        \includegraphics[width=\textwidth]{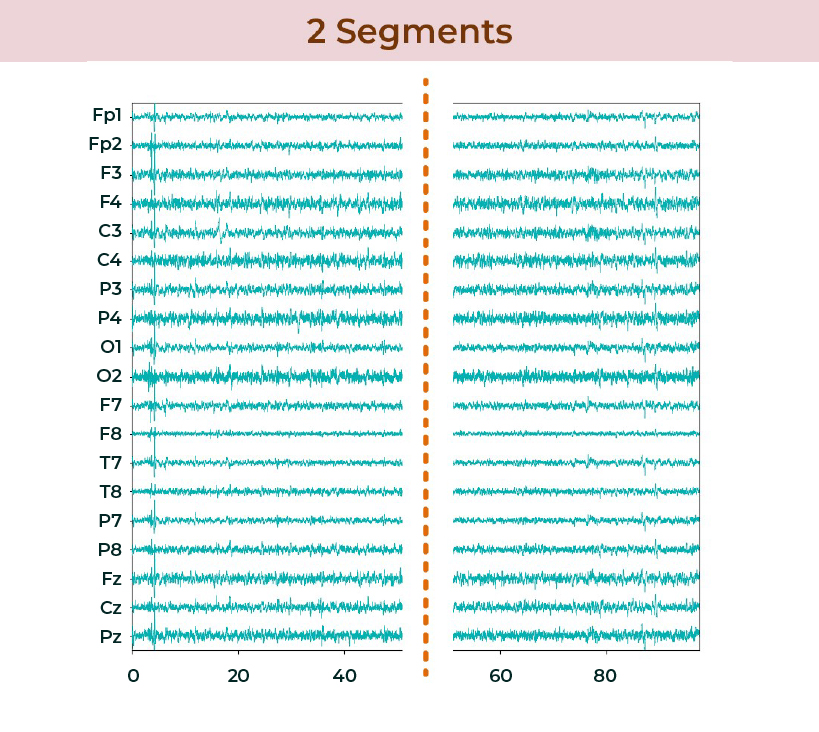}
    \caption{EEG recording divided into two equal segments.}
    \label{fig:2-segm}
\end{subfigure}
\begin{subfigure}{0.4\textwidth}
    \centering
        \includegraphics[width=\textwidth]{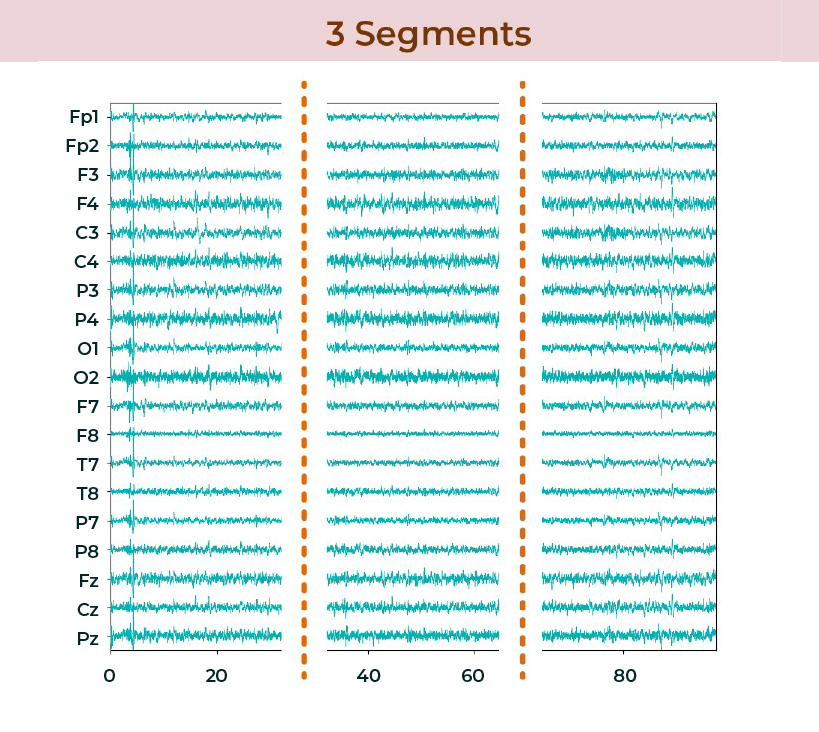}
    \caption{EEG recording divided into three equal segments.}
    \label{fig:3-segm}
\end{subfigure}
\begin{subfigure}{0.4\textwidth}
    \centering
        \includegraphics[width=\textwidth]{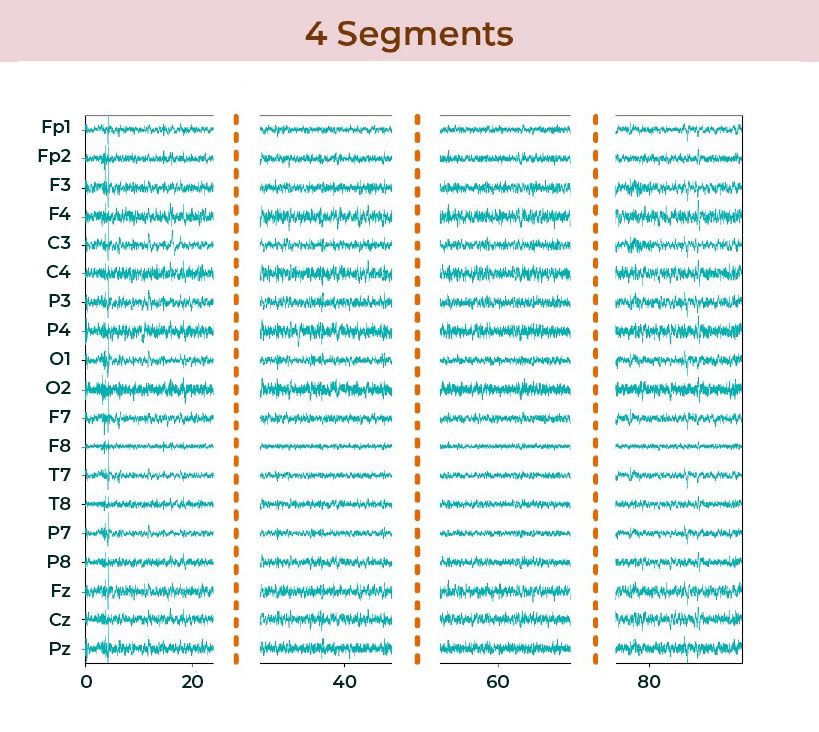}
    \caption{EEG recording divided into four equal segments.}
    \label{fig:4-segm}
\end{subfigure}
\begin{subfigure}{0.4\textwidth}
    \centering
        \includegraphics[width=\textwidth]{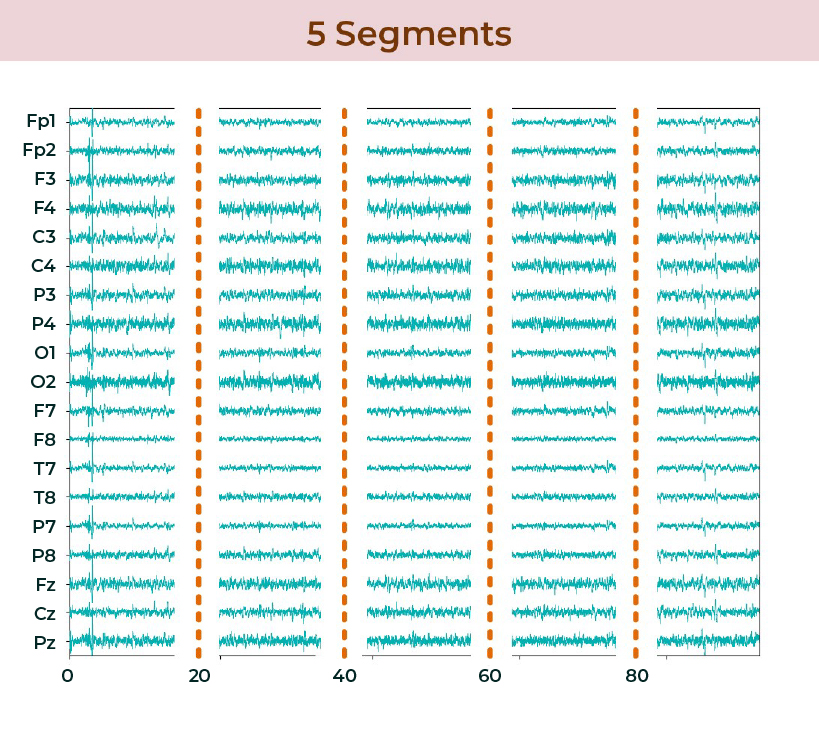}
    \caption{EEG recording divided into five equal segments.}
    \label{fig:5-segm}
\end{subfigure}    
\caption{Example of EEG recording segmentation showing a recording (channels vs. time duration) split into 2, 3, 4, and 5 segments.}
\label{fig:data-segm}
\end{figure}

Since each recording is now split into 2, 3, 4, 5, and 20 segments, each data matrix generates 35 new matrices ($1+2+3+4+5+20=35$ segments). The segmentation process is illustrated in Figure \ref{fig:data-segm}, showing how to divide one EEG recording into 2, 3, 4, and 5 segments. Segmentation into 20 parts follows the same procedure. This results in $4 \text{(preprocessings)} \times 35 \text{(segments)} = 140$ matrices per subject. The dimension of each matrix still depends on the subject's recording time, specifically $19 \times \frac{T_{i}}{j}$, where $j \in \{1,2,3,4,5,20\}$, $i \in \{0,120\}$, and $T_{i}$ is the duration of subject $i$'s experiment.

\subsection{Feature Extraction}\label{sec:feat-extraction}

\begin{figure}[ht]
    \centering
        \includegraphics[width=0.8\textwidth]{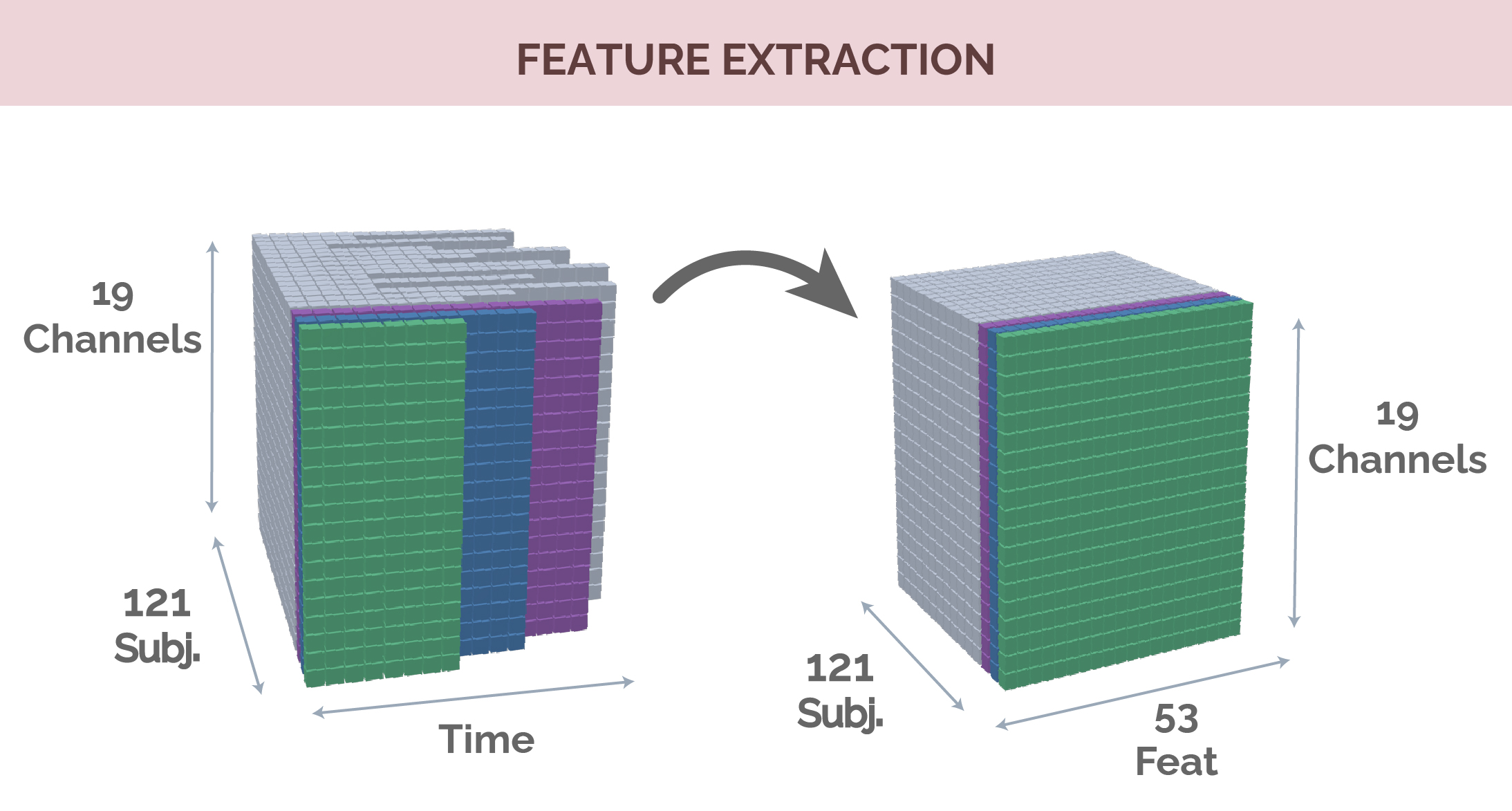}
    \caption{Visual explanation of the feature extraction process. Each coloured rectangle represents the data matrix of a subject.}
    \label{fig:feat-extract}
\end{figure}

After obtaining 140 matrices per subject, we proceed with the feature extraction process. Using the \textit{MNE-features} library \cite{schiratti2018ensemble} from MNE-Python, we extract 53 features, as detailed in Table \ref{tab:feat-ext}. These features are extracted independently for each channel; thus, for each matrix of dimensions $19 \times T_{i,j}$, we obtain 19 vectors, each containing 53 features (see Figure \ref{fig:feat-extract}).

In the final step, we generate the matrices to be used as input for the ML models, resulting in $4 \cdot 35 \cdot 19 = 2660$ matrices of size $121 \times 53$. It is important to note that the matrices are independent of the time spent conducting the experiment.

\begin{table}[ht]
\caption{Name and references of the 53 extracted features from EEG data.}
\label{tab:feat-ext}
\begin{tabular*}{\tblwidth}{CC}
 \toprule 
 \textbf{Extracted Features} & \textbf{Extracted Features} \\ 
 \midrule

 Mean & Hjorth complexity \cite{paivinen2005epileptic} \\

 Variance & Higuchi fractal dimension \cite{esteller2001comparison,paivinen2005epileptic} \\
 
 Standard deviation & Katz fractal dimension \cite{esteller2001comparison} \\

 Peak-to-peak amplitude & Number of zero-crossings \\
 
 Skewness & Line length \cite{esteller2001line} \\

 Kurtosis & Power Spectral Density intercept \cite{demanuele2007distinguishing, winkler2011automatic} \\
 
 Root-mean squared value & Power Spectral Density slope \cite{demanuele2007distinguishing, winkler2011automatic} \\

 Quantile & Power Spectral Density MSE \cite{demanuele2007distinguishing, winkler2011automatic} \\
 
 Hurst exponent \cite{qian2004hurst, devarajan2014eeg} & Power Spectral Density R2 \cite{demanuele2007distinguishing, winkler2011automatic} \\

 Approximate entropy \cite{richman2000physiological} & Spectral Entropy \cite{inouye1991quantification} \\
 
 Decorrelation time \cite{teixeira2011epilab} & Delta energy \cite{kharbouch2011algorithm} \\

 Delta power spectrum \cite{teixeira2011epilab} & Theta energy \cite{kharbouch2011algorithm} \\
 
 Theta power spectrum \cite{teixeira2011epilab} & Alpha energy \cite{kharbouch2011algorithm} \\

 Alpha power spectrum \cite{teixeira2011epilab} & Beta energy \cite{kharbouch2011algorithm} \\
 
 Beta power spectrum \cite{teixeira2011epilab} & Spectral Edge Frequency \cite{mormann2007seizure} \\

 Hjorth mobility from power spectrum \cite{mormann2007seizure, teixeira2011epilab} & Db4 Wavelet energy (x6 levels of decomposition) \cite{teixeira2011epilab} \\
 
 Hjorth complexity from power spectrum \cite{mormann2007seizure, teixeira2011epilab} & Teager-Kaiser energy (x14 levels of decomposition) \cite{badani2017detection} \\

 Hjorth mobility \cite{paivinen2005epileptic} & \\
 \bottomrule
\end{tabular*}
\end{table}

\subsection{Feature selection}\label{sec:feat-selection}

Feature selection is a key step in this type of experiment, as it helps identify the most significant features to improve model classification. We have opted for an inference-based feature selection approach, since the selection of statistically significant features has been shown to improve task prediction from EEG data \cite{degirmenci2023statistically}, reinforcing the importance of careful feature extraction for EEG-based classification. In the present study, a statistical test is applied to determine whether there are significant differences between the population means of ADHD and TD children for each feature. The following hypotheses are tested, where $\mu_{ADHD}$ and $\mu_{TD}$ represent the population means for ADHD and TD groups, respectively:

\begin{equation}
    \begin{array}{c}
        H_{0}: \mu_{ADHD} = \mu_{TD} \\
        H_{1}: \mu_{ADHD} \neq \mu_{TD}
    \end{array}
\end{equation}

In order to carry out this test, we must check whether both populations follow a normal distribution and whether there is homoscedasticity, i.e., if the variances are homogeneous. We set a significance level of $\alpha = 0.05$ for these tests. First, we assess normality using the D'Agostino and Pearson’s test \cite{d1971omnibus, d1973tests}, which combines skewness and kurtosis to produce an omnibus test of normality. If both populations (ADHD and TD children) follow a normal distribution, we then use Bartlett's test to check for homoscedasticity, which tests the null hypothesis that all input samples have equal variances. If at least one population does not follow a normal distribution, we apply Levene's test to assess homoscedasticity. 

Once we determine normality and homoscedasticity, we apply a statistical test to evaluate if the population means differ. If both populations are normal with equal variances, we use the classical t-test. If the populations are normal but the variances are not homogeneous, we use Welch's t-test. If at least one population is non-normal but the variances are homogeneous, we still use the classical t-test. However, if the populations are neither normal nor homogeneous, we cannot apply any test to reject the null hypothesis, and such features are excluded.

\begin{figure}[ht]
    \centering
    \includegraphics[width=0.8\textwidth]{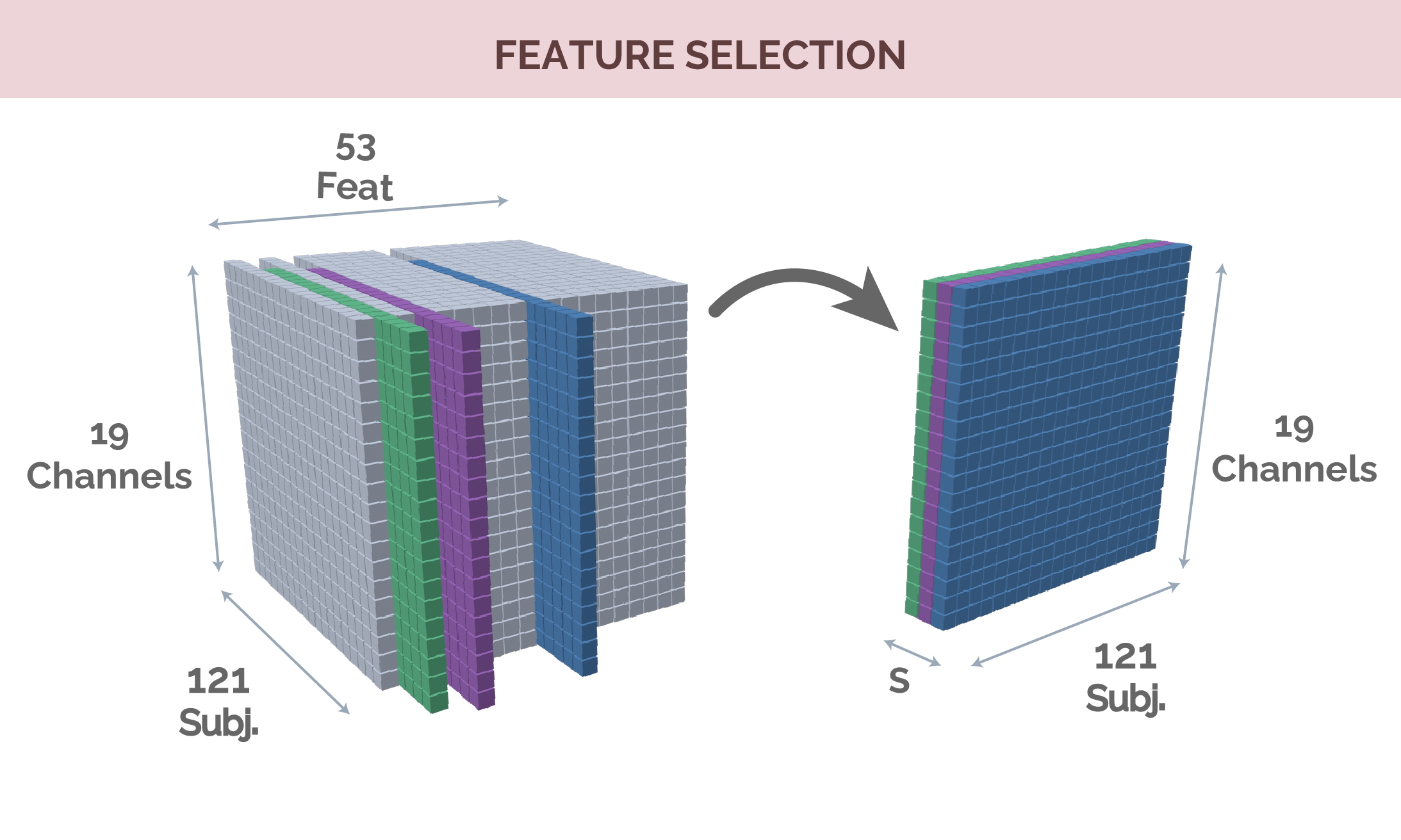}
    \caption{Visual explanation of the feature selection process, where $S$ is the number of the selected features.}
    \label{fig:feat-select}
\end{figure}

After performing these tests, we select the features/columns where there is evidence to reject the null hypothesis, i.e., features that show statistically significant differences between the population means (\textit{p-value} $< 0.05$). This process is illustrated in Figure \ref{fig:feat-select}. The procedure has been applied to each combination of cleaning type and segmentation data. For comparison, we have trained models using both the full dataset and the dataset containing only the selected features.

Once this step is completed, we add a label for each subject: \textit{1} for ADHD children and \textit{0} for TD children. This results in matrices of dimension $121 \times (S + 1)$, where $S$ is the number of selected features.

\subsection{Classification models and training process}\label{sec:clf-models}

Regarding the classification process, we need to run the model a large number of times, and we seek an explainable model due to the nature of the problem. The XGBoost model \cite{chen2016xgboost} addresses both needs. It offers fast and accurate solutions, as XGBoost is designed for high efficiency. It implements ML algorithms based on the Gradient Boosting framework and provides parallel tree boosting. Additionally, XGBoost calculates feature importance, based on the F score, once the boosted trees are built, offering explainability. The F score in XGBoost is a metric that represents the importance of a feature in the decision trees. It is the count of how many times a feature is used to split the data across all trees in the ensemble. The more frequently a feature is used to make decisions in the trees, the higher its F score, and thus, the more important that feature is considered in the model. It is important to say that the F score in the context of XGBoost refers to feature importance and not to the $f_{1}$-score used in classification.

To assess the robustness of the results, we also employ two widely used ML classifiers: Support Vector Machine (SVM) and K-Nearest Neighbors (KNN) \cite{deshmukh2024contributions, karimui2022adhd, sharma2023attention}. SVM is a supervised learning algorithm used for classification and regression, focusing on finding the optimal hyperplane to best separate data points into classes. It performs well in high-dimensional spaces and is robust against overfitting, particularly when the number of features exceeds the number of samples. KNN, on the other hand, is a simple and intuitive algorithm used for classification and regression. It classifies new data points based on the majority class among its K nearest neighbors in the feature space.

With regard to the data used to train these models, we use the matrices described in previous subsections of Section \ref{sec:matandmet}. Each matrix has different attributes, as the selected features vary, but they all have the same number of rows, corresponding to the 121 subjects. To generate a training matrix, we select one option from each of the following attributes:

\begin{itemize}
    \item \textbf{Cleaning process of the EEG signal}: No cleaning (raw), filtering technique, ASR algorithm, or ICA method.
    \item \textbf{Time segment of the EEG recording}: The whole recording, one half, one third, one fourth, one fifth, or one twentieth.
    \item \textbf{Channels selected}: One, two, or three channels.
    \item \textbf{Use of feature selection}: Yes or no.
    \item \textbf{Classification model}: XGBoost, SVM, or KNN.
\end{itemize}

We created all possible matrices using all combinations of the above attributes. Additionally, we applied a 5-fold cross-validation technique in each training process.

\section{Results}\label{results}

In this section, we present the results of the conducted experiments, displayed graphically to facilitate readability. Additionally, we highlight the most significant findings from each experiment. It is important to note that all experiments are fully reproducible, as we have used a public dataset, open-source software, and tools. The code is available on \href{https://gitlab.com/lucentia/refining-adhd-diagnosis-with-eeg-preprocesing-and-temporal-segmentation}{GitLab}.

First, we consolidated all the results obtained from training the ML models into a matrix (\textit{DataFrame} in \textit{Pandas}), and then grouped the matrix by its attributes to analyze the results, as described throughout this section. A sample of the complete matrix is shown in Table \ref{tab:df_results}. The full matrix contains 7,510,704 rows and 7 columns. It is important to note that this sample consists of real rows selected from the actual DataFrame. Each row contains the selected attributes used to form a training matrix, the ML model applied for classification, and its performance metrics. As shown, two columns are dedicated to model performance: the mean \textit{accuracy} and the \textit{standard error}, obtained after applying 5-fold cross-validation. The \textit{Cleaning technique} column refers to the type of preprocessing applied to the data (Raw, Filtered, ASR, or ICA), as explained in Section \ref{sec:prep}. The \textit{Segment} column indicates the segment of the recording used for training, whether it be the whole recording (1/1) or a fraction of the duration (1/2, 1/3, 1/4, 1/5, or 1/20), as detailed in Section \ref{sec:chunks}. The \textit{Channels} column shows the specific channel(s) used, which could range from one to three channels. The \textit{Classifier} column identifies the ML model applied, as explained in Section \ref{sec:clf-models}. Lastly, the \textit{Feature Selection} column indicates whether feature selection was used (Yes) or not (No), as discussed in Section \ref{sec:feat-selection}.

\begin{table*}[ht]
    \begin{center}
    \caption{Sample of the DataFrame containing all results: the attributes of the training matrix, the ML model used and the performance obtained in each training.}
    \label{tab:df_results} 
    \begin{tabular}{c c c c c c c}
        \toprule
        Accuracy & Standard error & Cleaning technique & Segment & Channels & Classifier & Feature Selection \\ 
        \midrule
        0.8847 & 0.0473 & Raw & 17/20 & Fp2-P3-T7 & XGBoost & Yes  \\ 
        0.8677 & 0.0756 & Raw & 15/20 & Fp2-O1-P7 & XGBoost & Yes \\
        0.8677 & 0.1109 & Raw & 3/3 & F4-Fz-Cz & XGBoost & Yes \\
        0.8610 & 0.0568 & ASR & 1/1 & P3-P4 & XGBoost & Yes \\
        0.8603 & 0.1194 & Raw & 17/20 & F4-P7 & XGBoost & No \\
        0.8523 & 0.1305 & ASR & 1/1 & P3-P4-Cz & XGBoost & Yes \\
        0.8437 & 0.0542 & ASR & 18/20 & F3-C3-P3 & XGBoost & Yes \\
        0.8433 & 0.0940 & ASR & 18/20 & Fp1-F3-C3 & XGBoost & Yes \\
        0.8433 & 0.0791 & Raw & 14/20 & P3 & XGBoost & No \\
        0.8357 & 0.1249 & ASR & 1/1 & P3-F8 & XGBoost & No \\
        0.8350 & 0.0244 & Filtered & 6/20 & Fp2-Cz-Pz & XGBoost & Yes \\
        0.8267 & 0.0903 & Raw & 19/20 & P7 & XGBoost & Yes \\
        0.8193 & 0.0222 & ICA & 3/3 & C4-P3-Fz & XGBoost & Yes \\
        0.8190 & 0.0184 & ICA & 1/4 & F4-O2-Fz & XGBoost & Yes \\
        0.8187 & 0.0250 & Raw & 17/20 & T8-P7 & XGBoost & No \\
        0.8180 & 0.1010 & Raw & 4/5 & Fp1-Fz-Cz & XGBoost & Yes \\
        0.8110 & 0.1194 & Filtered & 4/5 & Pz & KNN & Yes \\
        0.8107 & 0.1400 & ASR & 1/1 & F3-P3-Cz & XGBoost & Yes \\
        0.8103 & 0.0660 & ASR & 18/20 & F3 & XGBoost & No \\
        0.8103 & 0.0845 & Raw & 17/20 & P7 & SVM & Yes \\
        0.8090 & 0.1523 & ICA & 3/3 & P7-Fz-Cz & XGBoost & Yes \\
        \bottomrule
    \end{tabular}
    \end{center}
\end{table*}

Once all the results obtained from training the ML models are gathered into a single matrix, we can present the information more effectively. Since this work primarily focuses on the importance of EEG data preprocessing, we dedicate a subsection to results related to the type of preprocessing applied. We also consider the impact of EEG recording time when divided into segments, which is explored in a separate subsection. Additionally, insights regarding classifiers and feature selection methods are discussed, followed by a section focusing on channel-related outcomes. Finally, the results section concludes with a subsection on feature importance.

We primarily present the results using box plots. Each box plot displays the mean accuracy of the trained models on the vertical axis and different experimental variables on the horizontal axis. These box plots illustrate the distribution of model accuracies, grouped by the various columns of the DataFrame (Table \ref{tab:df_results}).

\subsection{Preprocessing results}\label{sec:prep-results}

\begin{figure}[ht]
    \centering
    \includegraphics[width=0.6\textwidth]{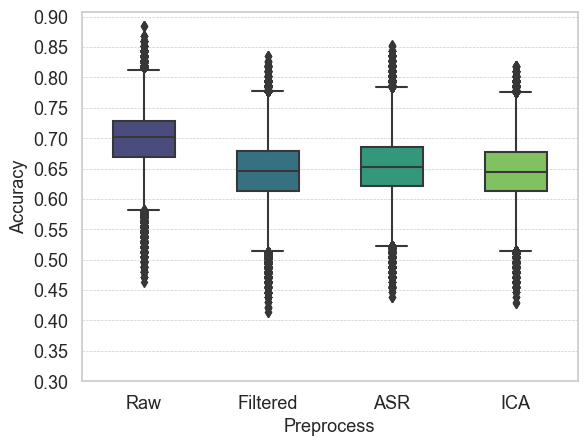}
    \caption{Box plot representation of model accuracies across quartiles, grouped by preprocessing technique.}
    \label{fig:preprocess}
\end{figure}

First, we present the overall performance of each preprocessing technique in Figure \ref{fig:preprocess}. When no preprocessing is applied (Raw), ML models achieve high accuracy levels. However, when noise is removed using the Filtered, ASR, or ICA techniques, overall accuracy decreases. A significant drop in accuracy can be observed between the Raw data (no noise removal) and the ICA-preprocessed data, which removes the most noise.

\begin{figure}[ht]
\begin{subfigure}{\textwidth}
    \centering
        \includegraphics[width=0.82\textwidth]{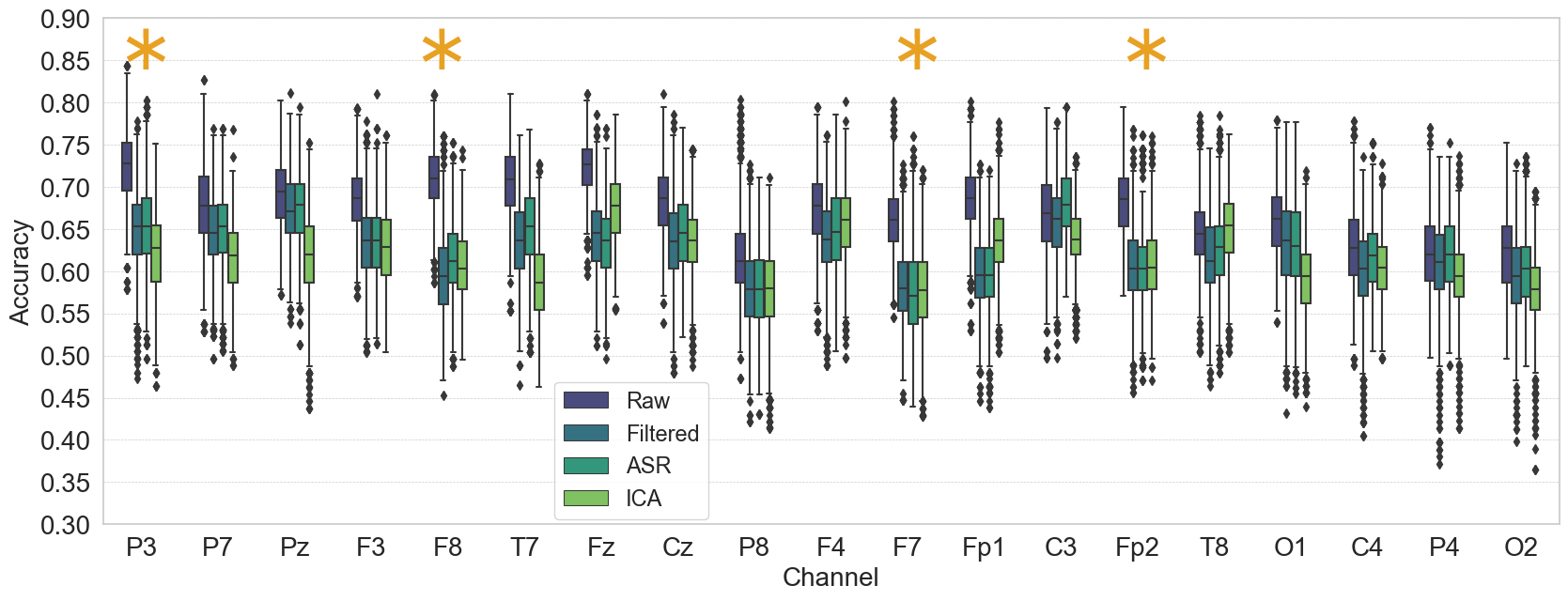}
    \caption{Distribution of accuracies divided by preprocessing technique and individual channels.}
    \label{fig:prep-1chan}
\end{subfigure}
\begin{subfigure}{\textwidth}
    \centering
        \includegraphics[width=0.82\textwidth]{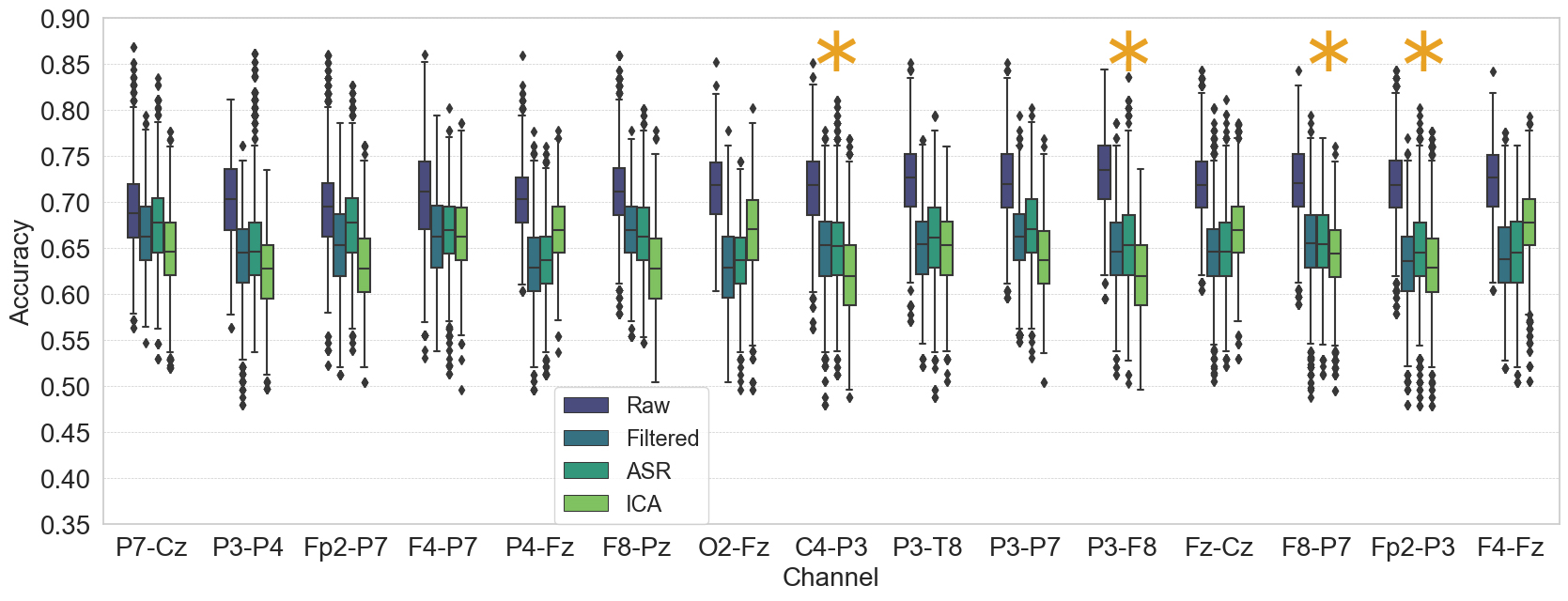}
    \caption{Distribution of accuracies divided by preprocessing technique and channel pairs.}
    \label{fig:prep-2chan}
\end{subfigure}
\begin{subfigure}{\textwidth}
    \centering
        \includegraphics[width=0.82\textwidth]{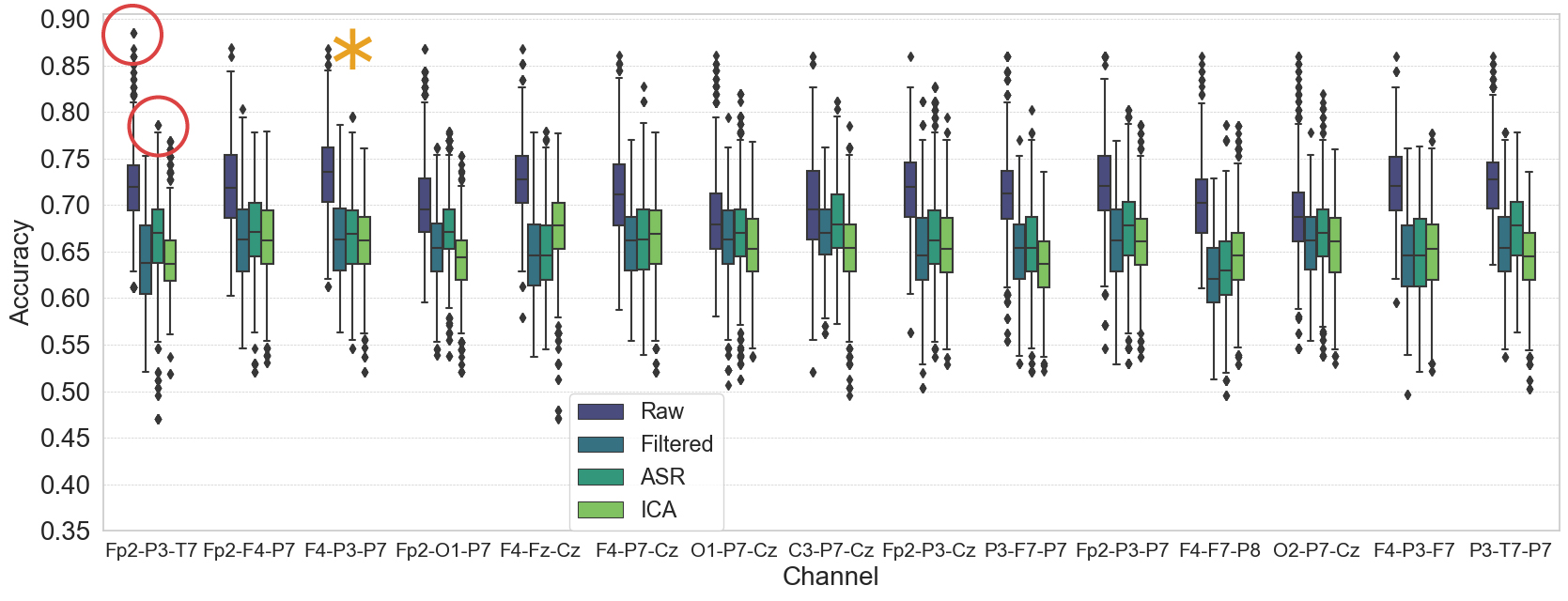}
    \caption{Distribution of accuracies divided by preprocessing technique and channel trios.}
    \label{fig:prep-3chan}
\end{subfigure}      
\caption{Distribution of accuracies divided by preprocessing technique and number of channels. The figures compare the model performance based on the number of channels used and highlight the statistically significant differences indicated by yellow stars between raw and preprocessed data across different channel combinations. These figures highlight the impact of preprocessing methods (filtered, ASR, and ICA) on model performance.}
\label{fig:prep-chan}
\end{figure}

When breaking down the results from Figure \ref{fig:preprocess} by channels, we obtain the results shown in Figure \ref{fig:prep-chan}, which displays the accuracy distribution of models by preprocessing technique and channel. Channels with statistically significant differences between \textit{Raw} and preprocessed data are marked with a yellow star. Figure \ref{fig:prep-1chan} illustrates the accuracy distribution by preprocessing technique for each channel, showing that channels P3, F8, F7, and Fp2 result in better model performance without preprocessing. 

Figure \ref{fig:prep-2chan} highlights the 15 most accurate channel pairs, where several (C4-P3, P3-F8, F8-P7, Fp2-P3) show statistically significant differences between using and not using preprocessing. 

Finally, Figure \ref{fig:prep-3chan} shows accuracy distributions for channel trios. While fewer trios show significant differences, larger differences are seen between maximum accuracies for \textit{Raw} data and preprocessed data. For instance, in one trio, the maximum accuracy drops from 0.88 with \textit{Raw} data to 0.78 with the \textit{ASR} algorithm, marked with red circles.

\subsection{Segmentation results}

In this section, we present results related to the segmentation of the EEG recordings, as defined in Section \ref{sec:chunks}. As observed in Section \ref{sec:prep-results}, not removing noise results in higher accuracy, but since we aim to focus on brain signal analysis, we exclude \textit{Raw} data to avoid noise influencing the results. Therefore, the results in this section are based on the \textit{Filtering}, \textit{ASR}, and \textit{ICA} preprocessing methods.

\begin{figure}[ht]
\centering
\begin{subfigure}{0.45\textwidth}
    \includegraphics[width=0.9\textwidth]{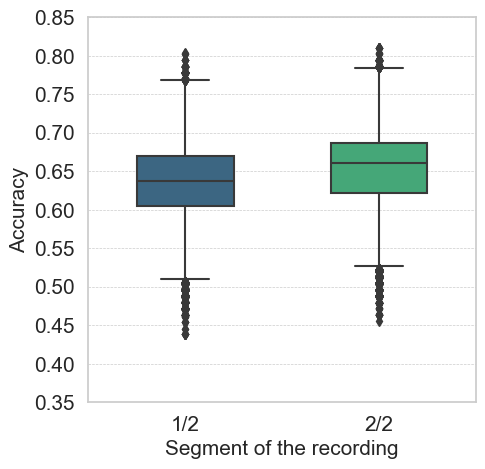}
    \caption{Accuracy distribution for models trained on recordings divided into 2 segments.}
    \label{fig:2chunks}
\end{subfigure}
\begin{subfigure}{0.45\textwidth}
    \includegraphics[width=0.9\textwidth]{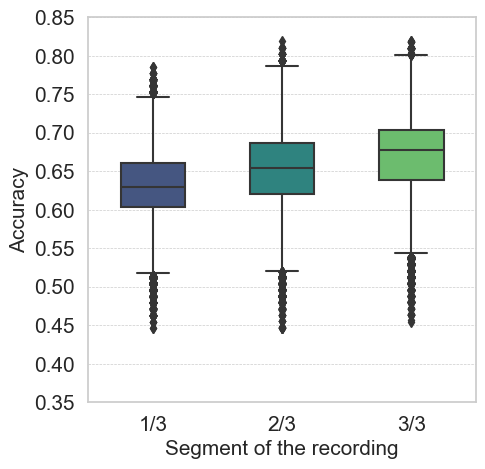}
    \caption{Accuracy distribution for models trained on recordings divided into 3 segments.}
    \label{fig:3chunks}
\end{subfigure}
\begin{subfigure}{0.45\textwidth}
    \includegraphics[width=0.9\textwidth]{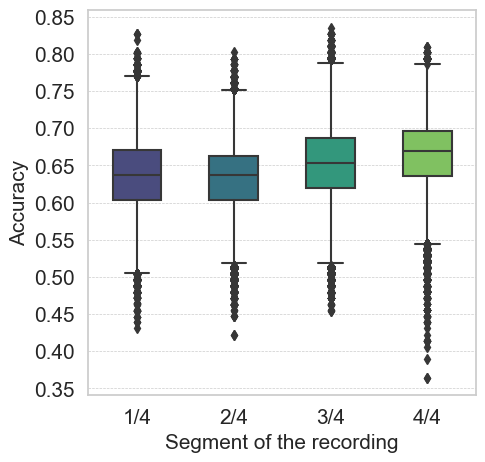}
    \caption{Accuracy distribution for models trained on recordings divided into 4 segments.}
    \label{fig:4chunks}
\end{subfigure}
\begin{subfigure}{0.45\textwidth}
    \includegraphics[width=0.9\textwidth]{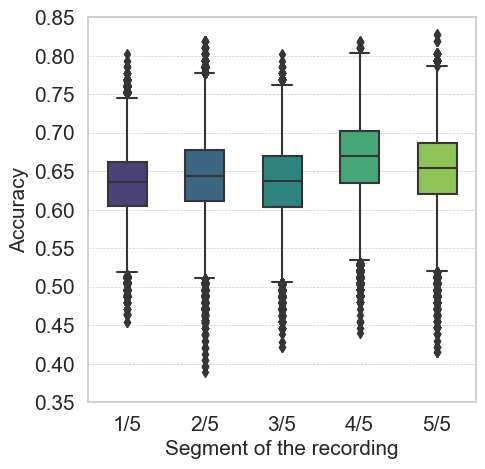}
    \caption{Accuracy distribution for models trained on recordings divided into 5 segments.}
    \label{fig:5chunks}
\end{subfigure}      
\caption{Box plots representing model accuracies based on the segmentation applied to the EEG recordings. Later segments tend to show improved model performance.}
\label{fig:chunks}
\end{figure}

We begin by analyzing the overall model performance based on the EEG segment used for training. Figure \ref{fig:chunks} presents box plots of accuracies when the EEG recordings are divided into 2 (Fig. \ref{fig:2chunks}), 3 (Fig. \ref{fig:3chunks}), 4 (Fig. \ref{fig:4chunks}), and 5 (Fig. \ref{fig:5chunks}) segments. The plot for 20 segments has been omitted due to its lack of relevant information. Across the four plots in Figure \ref{fig:chunks}, there is a noticeable improvement in model performance when later segments are used, particularly in the last two segments, as observed in Figures \ref{fig:3chunks}, \ref{fig:4chunks}, and \ref{fig:5chunks}.

\begin{figure}[htp]
\centering
\begin{subfigure}{\textwidth}
\centering
    \includegraphics[width=0.9\textwidth]{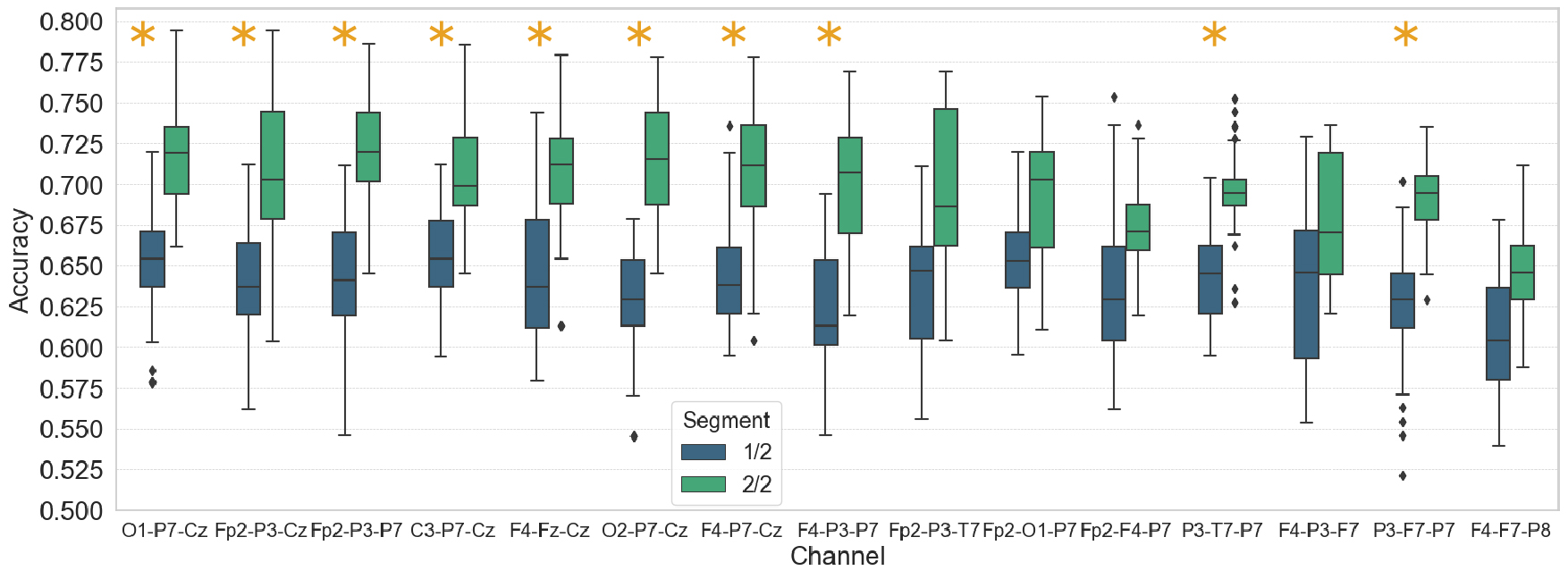}
    \caption{Accuracy distribution for models trained on recordings divided into 2 segments, according to the trio of channels selected.}
    \label{fig:2chunks-3chan}
\end{subfigure}
\begin{subfigure}{\textwidth}
\centering
    \includegraphics[width=0.9\textwidth]{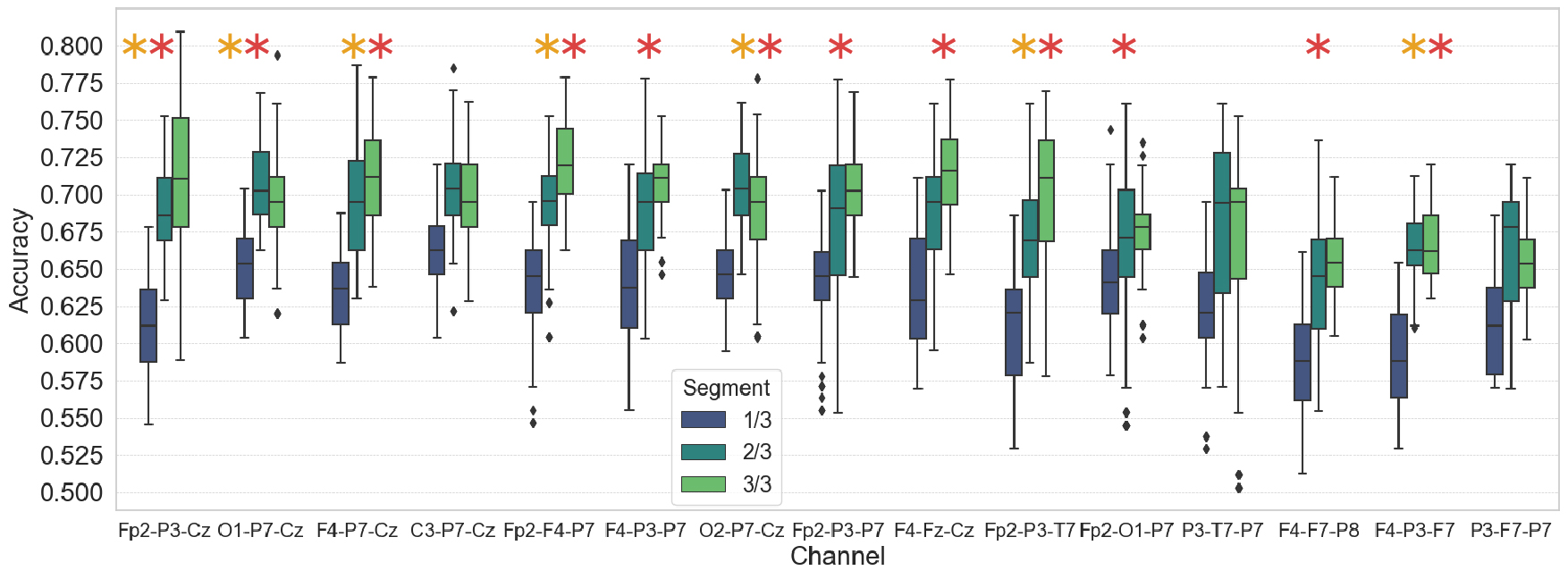}
    \caption{Accuracy distribution for models trained on recordings divided into 3 segments, according to the trio of channels selected.}
    \label{fig:3chunks-3chan}
\end{subfigure}
\caption{Box plots of model accuracies based on segmentation and trio of channels selected. In Figure \ref{fig:2chunks-3chan}, models trained on the first and second halves of the recordings that show significant differences are marked with a yellow star. In Figure \ref{fig:3chunks-3chan}, significant differences are observed between the first third and the remaining two-thirds (marked by a yellow star), and between the first and third thirds (marked by a red star).}

\label{fig:chunks-channels}
\end{figure}

Next, we analyzed the effect of segmentation based on the selected channels. We focused on data from one and two channels, but the differences among segments were not significant. However, interesting results emerged when examining trios of channels. Figure \ref{fig:chunks-channels} displays box plots of accuracies obtained from trios of channels, with recordings divided into 2 and 3 segments. Figure \ref{fig:2chunks-3chan} shows significant differences between the first and second halves of the recording, marked with a yellow star. In Figure \ref{fig:3chunks-3chan}, significant differences are observed between the first third and the remaining two-thirds, marked with a yellow star, while differences between the first and third thirds are indicated with a red star.

\subsection{Classifiers and feature selection results}

\begin{figure}[ht]
    \centering
    \includegraphics[width=0.6\textwidth]{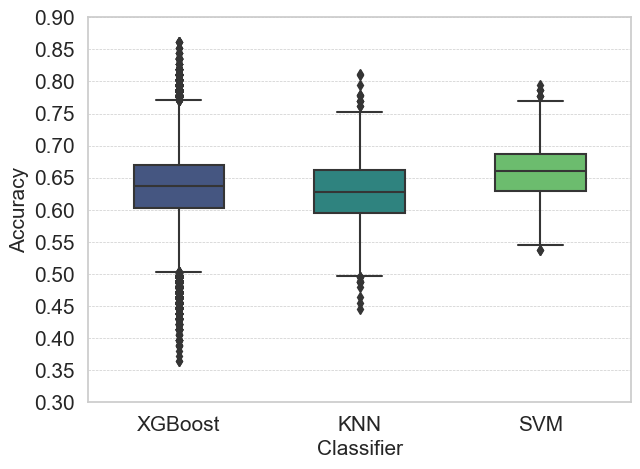}
    \caption{Distribution of accuracies divided by classifier.}
    \label{fig:classifiers}
\end{figure}

In this subsection, we present the differences between the classifiers described in Section \ref{sec:clf-models}. Figure \ref{fig:classifiers} displays box plots of model accuracies based on the ML classifier used. As before, results using \textit{Raw} data are excluded, and only models trained with one or two channels are considered. 

Two main points stand out in this figure. First, the SVM classifier has the largest interquartile range (IQR) and shorter whiskers, indicating less variability in its performance. Second, the XGBoost model achieves the highest overall accuracy but has a higher number of outliers, which is a consequence of its algorithmic structure.

Since the experiment with trios of channels was computationally intensive and time-consuming, we decided to run it only using XGBoost, as it is faster than SVM and KNN, and provided the highest results with one and two channels.

\begin{figure}[ht]
\centering
\begin{subfigure}{0.4\textwidth}
    \centering
    \includegraphics[width=\textwidth]{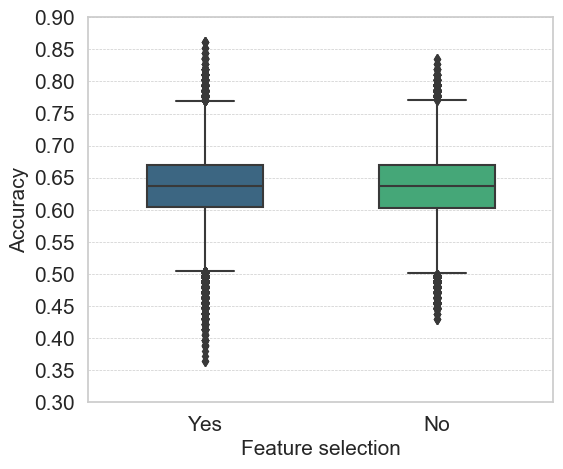}
    \caption{Box plot comparing model accuracies trained with selected features versus all features.}
    \label{fig:picked_features}
\end{subfigure}
\begin{subfigure}{0.4\textwidth}
    \centering
    \includegraphics[width=\textwidth]{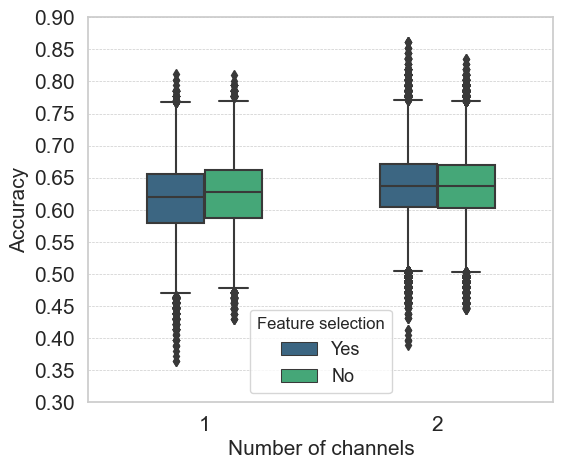}
    \caption{Box plots of model accuracies based on Feature selection and number of channels used.}
    \label{fig:picked_features_n_chan}
\end{subfigure}
\caption{Box plots illustrating model accuracies with feature selection versus using all features. (a) Shows the overall accuracy comparison, and (b) highlights the accuracy variation based on the number of channels used.}
\label{fig:2picked-features}
\end{figure}

Regarding feature selection, Figure \ref{fig:2picked-features} shows the impact of applying our feature selection method compared to not using it, as explained in Section \ref{sec:feat-selection}. In Figure \ref{fig:picked_features}, the similar IQR between both box plots indicates that removing features does not affect the results, which supports our method. Additionally, the highest accuracy is achieved when selecting features that show statistically significant differences between ADHD and TD children.

Breaking down the accuracy further, Figure \ref{fig:picked_features_n_chan} highlights the model performance based on the number of channels used. When using a single channel, there is no difference in maximum accuracy with or without feature selection, but the IQR decreases when feature selection is applied. However, when adding a second channel, the IQRs remain similar, and the highest accuracy is achieved with feature selection.

For the experiment involving three channels, we again used only features that displayed statistically significant differences between ADHD and TD children. This reduced the amount of data fed to the model and, consequently, the overall training time.

\subsection{Channels results}

\begin{figure}[ht]
    \centering
    \includegraphics[width=\textwidth]{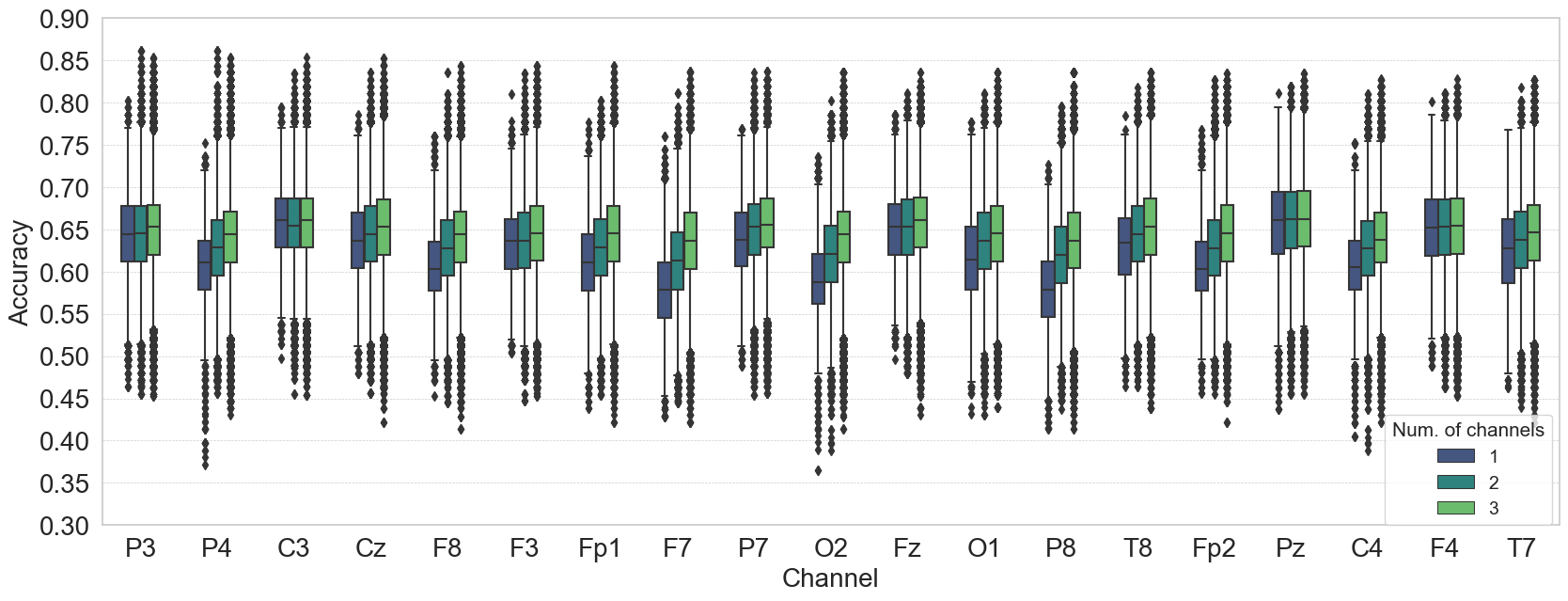}
    \caption{Box plots illustrating the accuracy of models as a function of the number of channels used. The distribution of accuracies is shown for models trained with a fixed channel, and then compared when adding one and two additional channels.}
    \label{fig:channels-n-chan}
\end{figure}

This section highlights the most relevant channels for classifying ADHD and TD children. When we break down the overall accuracy by channel according to the number of channels used, we obtain Figure \ref{fig:channels-n-chan}. The figure reveals two distinct groups of channels. The first group maintains a relatively constant IQR, regardless of the number of channels involved, and includes channels P3, C3, Fz, Pz, and F4. The second group shows a gradual increase in IQR as additional channels are included, consisting of P4, F8, Fp1, F7, O2, O1, P8, Fp2, and C4. Channels such as Cz, F3, P7, T8, and T7 do not exhibit a clear preference for either group.

We can infer that the channels in the second group (with increasing IQR as the number of channels increases) do not provide highly relevant information to the classifier on their own. However, when combined with other channels, they become useful. Conversely, channels in the first group (with relatively constant IQR) provide valuable information individually.

In this section, we can also present channel results differently. Since we know the location of each channel on the scalp, we can display a head diagram along with the corresponding values for better visualization.

\begin{figure}[ht]
\centering
    \begin{subfigure}{.4\linewidth}
        \centering
            \includegraphics[width=\textwidth]{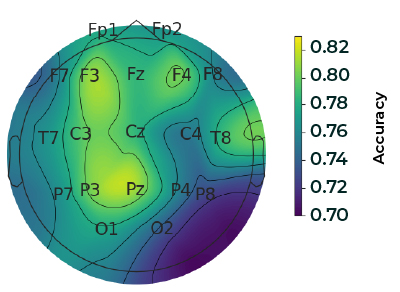}
        \caption{Topographic representation of accuracies for models trained using data from individual EEG channels.}
        \label{fig:head-1chan}
    \end{subfigure}%
    \hspace{0.3cm}\begin{subfigure}{.4\linewidth}
        \centering
            \includegraphics[width=\textwidth]{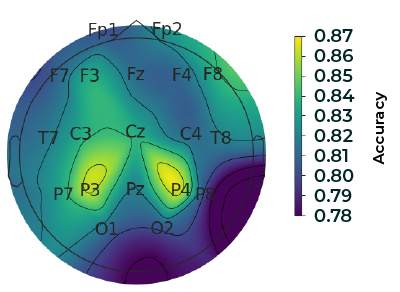}
        \caption{Topographic representation of accuracies for models trained using data from pairs of EEG channels.}
        \label{fig:head-2chan}
    \end{subfigure}\\[1ex]
    \hspace{0.2cm}\begin{subfigure}{\linewidth}
        \centering
            \includegraphics[width=0.4\textwidth]{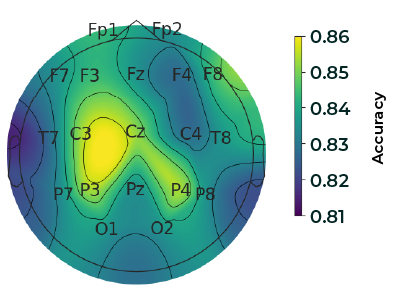}
        \caption{Topographic representation of accuracies for models trained using data from three EEG channels.}
        \label{fig:head-3chan}
    \end{subfigure}
    \caption{Topographic maps illustrating the accuracies obtained from models trained with different configurations: using one, two, and three EEG channels. The color gradient indicates classification performance, with higher accuracy highlighted in yellow and lower accuracy in purple.}
    \label{fig:head-chan}
\end{figure}

Figure \ref{fig:head-chan} shows that the most significant channels are located on the left side of the head and near the center, particularly around the Cz channel. When using only one channel to train the models, Figure \ref{fig:head-1chan} indicates that F3, P3, and Pz achieve the highest accuracies, followed by C3, Cz, F4, and T8. Figure \ref{fig:head-2chan} illustrates that P3 and P4 provide the best results for pairs of channels, with F3, C3, Cz, P7, and F8 also performing well. Lastly, with three channels, Figure \ref{fig:head-3chan} highlights C3, P3, Cz, and P4 as the most relevant. Overall, the least influential areas are on the rear right and left peripheral sides. These figures were generated using the YASSA package \cite{vallat2021open} in Python.

\subsection{Features results}

In this section, we present the feature importance of the best-performing models. We selected the top three models with the highest accuracy, trained using one, two, and three channels. It is important to remember that we refer to F score as feature importance in XGBoost context and not to the $f_{1}$-score used in classification. 

These three models were trained using XGBoost with \textit{objective}=\textit{binary:logistic}, \textit{number of rounds}=100, and \textit{early stopping rounds}=50, applying feature selection to include only features showing statistically significant differences between the two groups, as detailed in Section \ref{sec:feat-selection}. We performed a stratified train/test split, using 80\% of subjects for training and 20\% for testing. Given the large number of features, we display only the top 15 most important features for clarity.

\begin{figure}[ht]
    \centering
    \includegraphics[width=0.7\textwidth]{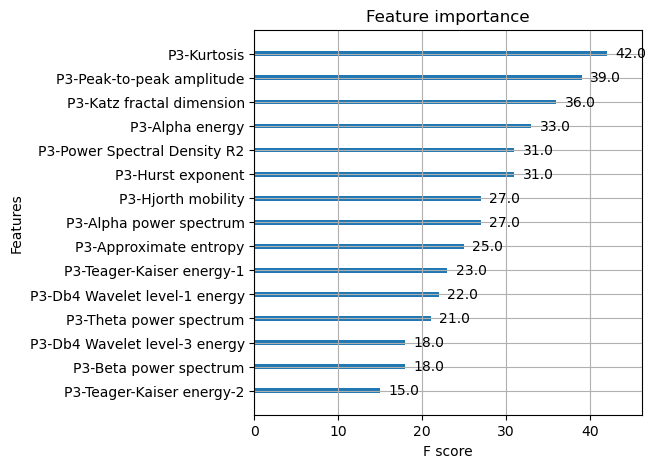}
    \caption{Feature importance of the best-performing model using one channel (P3). The top 15 most important features are displayed, as identified by the XGBoost model with feature selection applied.}
    \label{fig:feature-importance-p3}
\end{figure}

The first model presented is trained using data from the P3 channel, preprocessed with the ASR algorithm, and includes information from the entire EEG recording without segmentation. The XGBoost hyperparameters used are \textit{max depth}=12, \textit{eta}=0.1, and \textit{gamma}=0. Figure \ref{fig:feature-importance-p3} shows the feature importance of this model. Notably, the three most important features (Kurtosis, Peak-to-peak amplitude, and Katz fractal dimension) are related to the signal's shape. The Alpha frequency band is the most significant, followed by Theta and Beta. Additionally, non-linear features like Katz fractal dimension, Hurst exponent, Hjorth mobility, Approximate entropy, and Teager-Kaiser energies are prominent. The Daubechies-4 wavelet energies at decomposition levels 1 and 3 also rank among the top fifteen features.

\begin{figure}[ht]
    \centering
    \includegraphics[width=0.7\textwidth]{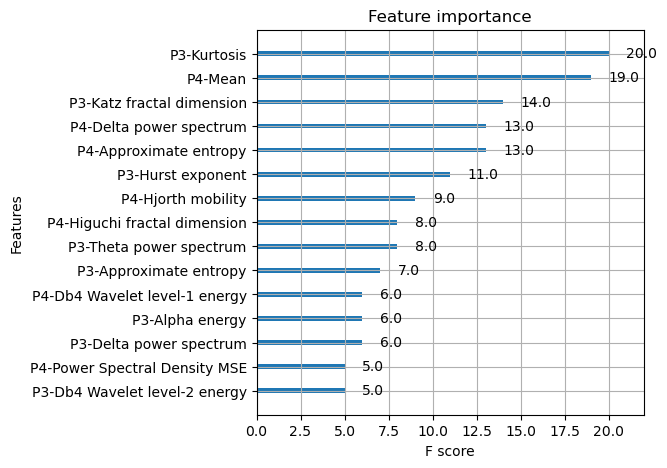}
    \caption{Feature importance of the best-performing model using two channels (P3 and P4). The top 15 most important features are displayed, as identified by the XGBoost model with feature selection applied.}
    \label{fig:feature-importance-p3p4}
\end{figure}

The second model presented achieved the highest accuracy using data from two channels (P3 and P4), with ASR preprocessing and the entire EEG recording, without segmentation. The optimal hyperparameters for this model were \textit{max depth}=6, \textit{eta}=0.1, and \textit{gamma}=1. Figure \ref{fig:feature-importance-p3p4} shows the feature importance for this model. Once again, Kurtosis (P3 channel) ranks first, and Katz fractal dimension (P3 channel) is third. Delta band features provide the most useful information, followed by Theta and Alpha bands. Non-linear features such as Katz and Higuchi fractal dimensions, Approximate entropy, Hurst exponent, and Hjorth mobility are significant. Additionally, two Daubechies-4 wavelet energies appear among the top features, with level 1 decomposition for the P4 channel and level 2 for P3.

\begin{figure}[htp]
    \centering
    \includegraphics[width=0.7\textwidth]{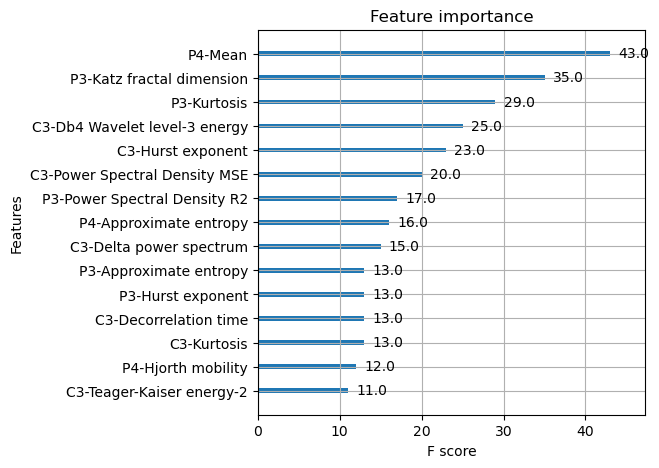}
    \caption{Feature importance of the best-performing model using three channels (C3, P3 and P4). The top 15 most important features are displayed, as identified by the XGBoost model with feature selection applied.}
    \label{fig:feature-importance-c3p3p4}
\end{figure}

Finally, we present the best three-channels model. It is trained with the data of C3, P3 and P4 channels, using the ASR preprocessing and the whole EEG recording. We set the hyperparameters of the XGBoost algorithm as: \textit{max depth}=3, \textit{eta}=0.1 and \textit{gamma}=0. As we can see in Figure \ref{fig:feature-importance-c3p3p4}, the first three positions are the same as the P3-P4 model (Figure \ref{fig:feature-importance-p3p4}), but in a different order. We also have non-linear features like Katz fractal dimension (P3 channel), Hurst exponent (C3 and P3 channels), Approximate entropy (P4 and P3 channels) and Hjorth mobility (P4 channel). In this case, there only is one feature related with frequency bands -Delta power spectrum-, and one Daubechies-4 wavelet energy feature with level 3 of decomposition.

\section{Discussion}\label{disc}

In this section, we highlight and discuss the most important results. First, we emphasize the importance of preprocessing, as demonstrated in Figures \ref{fig:preprocess} and \ref{fig:prep-chan}. Preprocessing is crucial when working with EEG signals from children with ADHD. Without preprocessing, i.e., without removing noise and artifacts, ML models produce artificially high accuracy. This may be because ADHD subjects exhibit symptoms like restlessness or difficulty concentrating, which can produce a higher number of artefacts and may result in models relying on inappropriate criteria. To better understand ADHD, it is essential to focus on brain-related signals, since not all ADHD patients have the same symptomatology.

We observed that several channels in this dataset, including P3, F8, T7, Fz, F7, and Fp2, collect significant amounts of noise, as shown in Figure \ref{fig:prep-1chan}. This is a common challenge when working with EEG data, as scalp sensors pick up a wide range of signals, including muscle movements, eye blinks, and heartbeats. The key to addressing this issue is implementing an effective preprocessing pipeline that removes unwanted noise while preserving brain data, ensuring that ML model results are not artificially inflated.

This is the first study to split EEG recordings into segments to investigate differences over time. We found that as the recording progresses, the models' decision-making becomes clearer. As shown in Figure \ref{fig:chunks-channels}, much better results are achieved when using the second half (Figure \ref{fig:2chunks-3chan}) or the second and third thirds of the EEG recordings (Figure \ref{fig:3chunks-3chan}). This could be due to ADHD symptomatology, where subjects may lose focus or become distracted as the test lengthens, more so than TD subjects.

It is noteworthy that the P3 channel proves highly useful for ML models, providing the most valuable data when using a single sensor and appearing in the best combinations with two and three channels. Another important channel is P4, though in a different way. While P4 does not offer much information on its own, it significantly enhances model performance when combined with other channels. These findings are consistent with other studies \cite{talebi2022investigating, gu2023detection, ekhlasi2021direction}, which also highlight the left brain as critical for identifying ADHD. Additionally, these results align with \cite{sanchis2024novel}, \cite{khare2023explainable}, and \cite{kasim2023identification}, which also identify P3 and P4 as key channels. As shown in Figure \ref{fig:head-chan}, the most influential channels in this study are located around the Cz channel, which \cite{abedinzadeh2023potential} also considers crucial in their research.

Regarding the importance of features identified by the XGBoost model, the most significant ones are consistently highlighted in Figures \ref{fig:feature-importance-p3}, \ref{fig:feature-importance-p3p4}, and \ref{fig:feature-importance-c3p3p4}. These key features include Kurtosis, Katz fractal dimension, Delta, Theta, and Alpha power spectrums, Hurst exponent, Hjorth mobility, Approximate entropy, Daubechies-4 wavelet energy, and Power Spectral Density. This aligns with the findings of \cite{gu2023detection}, \cite{khare2022vhers}, and \cite{taghibeyglou2022detection}, who also highlight Delta and Theta waves as important. Similarly, \cite{abbas2021effective}, \cite{khare2023explainable}, and \cite{lin2023measurement} identify entropy as a key feature in ADHD detection.

It is important to note that all the techniques used in these experiments are relatively straightforward. We extracted features using a public API, applied statistical tests for feature selection, and employed three widely-used ML models: SVM, KNN, and XGBoost for classification. The simplicity of these methods highlights the significance of our results, achieving 86\% accuracy with EEG recordings from just three channels.

However, this work has some limitations. Due to the scarcity of public ADHD databases with EEG data in children, we were only able to test our results on a single dataset, limiting the generalizability of our findings. Additionally, while the extracted features guide us toward understanding the disorder, they may not be descriptive enough for specialists to interpret easily, requiring further research in this area. Another limitation is the number of channels used; although 19 channels were available, we could only train up to 3 due to computational constraints, as mentioned in our previous work \cite{sanchis2024novel}.

\section{Conclusion}\label{concl}

This study marks a crucial step in ADHD research, proving the necessity of thorough preprocessing to ensure reliable and unbiased results from EEG signals in children. Moreover, we were the first to show the significant impact of using later segments of EEG recordings, where accuracy improves substantially, highlighting potential fatigue effects in ADHD patients. The results establish P3, P3-P4, and C3-P3-P4 as the most effective channels for diagnosis, with accuracy reaching up to 86.10\%. Furthermore, we identified key features such as Kurtosis, Katz fractal dimension, Delta, Theta, and Alpha power spectrums, Hurst exponent, Hjorth mobility, Approximate entropy, Daubechies-4 wavelet energy, and Power Spectral Density. as critical for classification.

Our findings underscore the need for more extensive EEG databases focusing on children with ADHD, which will enable further research and aid in early, accurate diagnosis. This work also opens new possibilities for studying cognitive fatigue and concentration challenges in ADHD patients, particularly as test length increases. By isolating the most relevant channels and features, this research provides a solid foundation for future investigations and advances in ADHD diagnosis through EEG.

\section*{Acknowledgements}
This work has been co-funded by the BALLADEER (PROMETEO/2021/088) project, a Big Data analytical platform for the diagnosis and treatment of Attention Deficit Hyperactivity Disorder (ADHD) featuring extended reality, funded by the Conselleria de Innovación, Universidades, Ciencia y Sociedad Digital (Generalitat Valenciana) and the AETHER-UA project (PID2020-112540RB-C43), a smart data holistic approach for context-aware data analytics: smarter machine learning for business modelling and analytics, funded by Spanish Ministry of Science and Innovation. Sandra García-Ponsoda holds a predoctoral contract granted by \textit{ValgrAI - Valencian Graduate School and Research Network of Artificial Intelligence} and the \textit{Generalitat Valenciana}, and co-funded by the \textit{European Union}. In addition, this work is part of the TSI-100927-2023-6 Project, funded by the Recovery, Transformation and Resilience Plan from the European Union Next Generation through the Ministry for Digital Transformation and the Civil Service.

\section*{Declaration of competing interest}
The authors declare that they have no conflict of interest.

\bibliographystyle{elsarticle-num} 

\bibliography{cas-refs}

\begin{thebibliography}{10}
\expandafter\ifx\csname url\endcsname\relax
  \def\url#1{\texttt{#1}}\fi
\expandafter\ifx\csname urlprefix\endcsname\relax\def\urlprefix{URL }\fi
\expandafter\ifx\csname href\endcsname\relax
  \def\href#1#2{#2} \def\path#1{#1}\fi

\bibitem{salari2023global}
N.~Salari, H.~Ghasemi, N.~Abdoli, A.~Rahmani, M.~H. Shiri, A.~H. Hashemian, H.~Akbari, M.~Mohammadi, The global prevalence of adhd in children and adolescents: a systematic review and meta-analysis, Italian Journal of Pediatrics 49~(1) (2023) 48.

\bibitem{song2021prevalence}
P.~Song, M.~Zha, Q.~Yang, Y.~Zhang, X.~Li, I.~Rudan, The prevalence of adult attention-deficit hyperactivity disorder: A global systematic review and meta-analysis, Journal of global health 11 (2021).

\bibitem{diagnostic1994statistical}
A.~Diagnostic, statistical manual of mental disorders. 4th edn american psychiatric association, Washington, DC (1994).

\bibitem{harpin2005effect}
V.~A. Harpin, The effect of adhd on the life of an individual, their family, and community from preschool to adult life, Archives of disease in childhood 90~(suppl 1) (2005) i2--i7.

\bibitem{alma9918660570001341}
A.~P. Association, Diagnostic and statistical manual of mental disorders : DSM-5™., 5th Edition, American Psychiatric Publishing, a division of American Psychiatric Association, Washington, DC ;, 2013 - 2013.

\bibitem{lola2019attention}
H.~M. Lola, H.~Belete, A.~Gebeyehu, A.~Zerihun, S.~Yimer, K.~Leta, et~al., Attention deficit hyperactivity disorder (adhd) among children aged 6 to 17 years old living in girja district, rural ethiopia, Behavioural neurology 2019 (2019).

\bibitem{el2023adult}
S.~El~Archi, S.~Barrault, M.~Garcia, S.~Branger, D.~Maug{\'e}, N.~Ballon, P.~Brunault, Adult adhd diagnosis, symptoms of impulsivity, and emotional dysregulation in a clinical sample of outpatients consulting for a behavioral addiction, Journal of Attention Disorders 27~(7) (2023) 731--742.

\bibitem{marshall2021diagnosing}
P.~Marshall, J.~Hoelzle, M.~Nikolas, Diagnosing attention-deficit/hyperactivity disorder (adhd) in young adults: A qualitative review of the utility of assessment measures and recommendations for improving the diagnostic process, The Clinical Neuropsychologist 35~(1) (2021) 165--198.

\bibitem{karavallil2023alterations}
S.~Karavallil~Achuthan, D.~Stavrinos, H.~B. Holm, S.~A. Anteraper, R.~K. Kana, Alterations of functional connectivity in autism and attention-deficit/hyperactivity disorder revealed by multi-voxel pattern analysis, Brain Connectivity 13~(9) (2023) 528--540.

\bibitem{lohani2023adhd}
D.~C. Lohani, B.~Rana, Adhd diagnosis using structural brain mri and personal characteristic data with machine learning framework, Psychiatry Research: Neuroimaging 334 (2023) 111689.

\bibitem{serrallach2022neuromorphological}
B.~L. Serrallach, C.~Gro{\ss}, M.~Christiner, S.~Wildermuth, P.~Schneider, Neuromorphological and neurofunctional correlates of adhd and add in the auditory cortex of adults, Frontiers in Neuroscience 16 (2022) 850529.

\bibitem{millevert2023resting}
C.~Millevert, N.~Vidas-Guscic, L.~Vanherp, E.~Jonckers, M.~Verhoye, S.~Staelens, D.~Bertoglio, S.~Weckhuysen, Resting-state functional mri and pet imaging as noninvasive tools to study (ab) normal neurodevelopment in humans and rodents, Journal of Neuroscience 43~(49) (2023) 8275--8293.

\bibitem{hassan2024convolutional}
U.~Hassan, A.~Singhal, Convolutional neural network framework for eeg-based adhd diagnosis in children, Health Information Science and Systems 12~(1) (2024) 44.

\bibitem{latifi2024siamese}
B.~Latifi, A.~Amini, A.~M. Nasrabadi, Siamese based deep neural network for adhd detection using eeg signal, Computers in Biology and Medicine 182 (2024) 109092.

\bibitem{jahani2024efficient}
H.~Jahani, A.~A. Safaei, Efficient deep learning approach for diagnosis of attention-deficit/hyperactivity disorder in children based on eeg signals, Cognitive Computation (2024) 1--16.

\bibitem{degirmenci2024eeg}
M.~Degirmenci, Y.~K. Yuce, M.~Perc, Y.~Isler, Eeg-based finger movement classification with intrinsic time-scale decomposition, Frontiers in Human Neuroscience 18 (2024) 1362135.

\bibitem{zhang2024review}
X.~Zhang, X.~Zhang, Q.~Huang, Y.~Lv, F.~Chen, A review of automated sleep stage based on eeg signals, Biocybernetics and Biomedical Engineering (2024).

\bibitem{altaheri2023deep}
H.~Altaheri, G.~Muhammad, M.~Alsulaiman, S.~U. Amin, G.~A. Altuwaijri, W.~Abdul, M.~A. Bencherif, M.~Faisal, Deep learning techniques for classification of electroencephalogram (eeg) motor imagery (mi) signals: A review, Neural Computing and Applications 35~(20) (2023) 14681--14722.

\bibitem{wolpaw2012brain}
J.~Wolpaw, E.~Wolpaw, \href{https://books.google.es/books?id=0dST2Lg4KVYC}{Brain-Computer Interfaces: Principles and Practice}, Oxford University Press, USA, 2012.
\newline\urlprefix\url{https://books.google.es/books?id=0dST2Lg4KVYC}

\bibitem{jurcak200710}
V.~Jurcak, D.~Tsuzuki, I.~Dan, 10/20, 10/10, and 10/5 systems revisited: their validity as relative head-surface-based positioning systems, Neuroimage 34~(4) (2007) 1600--1611.

\bibitem{bigdely2015prep}
N.~Bigdely-Shamlo, T.~Mullen, C.~Kothe, K.-M. Su, K.~A. Robbins, The prep pipeline: standardized preprocessing for large-scale eeg analysis, Frontiers in neuroinformatics 9 (2015) 16.

\bibitem{nolan2010faster}
H.~Nolan, R.~Whelan, R.~B. Reilly, Faster: fully automated statistical thresholding for eeg artifact rejection, Journal of neuroscience methods 192~(1) (2010) 152--162.

\bibitem{mumtaz2021review}
W.~Mumtaz, S.~Rasheed, A.~Irfan, Review of challenges associated with the eeg artifact removal methods, Biomedical Signal Processing and Control 68 (2021) 102741.

\bibitem{alsharif2024diagnosis}
N.~Alsharif, M.~H. Al-Adhaileh, M.~Al-Yaari, Diagnosis of attention deficit hyperactivity disorder: A deep learning approach, AIMS Mathematics 9~(5) (2024) 10580--10608.

\bibitem{cura2024detection}
O.~K. Cura, A.~Akan, S.~K. Atli, Detection of attention deficit hyperactivity disorder based on eeg feature maps and deep learning, Biocybernetics and Biomedical Engineering 44~(3) (2024) 450--460.

\bibitem{sharma2023classification}
Y.~Sharma, B.~K. Singh, Classification of children with attention-deficit hyperactivity disorder using wigner-ville time-frequency and deep expeegnetwork feature-based computational models, IEEE Transactions on Medical Robotics and Bionics (2023).

\bibitem{loh2022automated}
H.~W. Loh, C.~P. Ooi, P.~D. Barua, E.~E. Palmer, F.~Molinari, U.~R. Acharya, Automated detection of adhd: Current trends and future perspective, Computers in Biology and Medicine 146 (2022) 105525.

\bibitem{nasrabadi2020eeg}
A.~M. Nasrabadi, A.~Allahverdy, M.~Samavati, M.~R. Mohammadi, Eeg data for adhd/control children, IEEE Dataport (2020).

\bibitem{abedinzadeh2023potential}
F.~Abedinzadeh~Torghabeh, S.~A. Hosseini, Y.~Modaresnia, Potential biomarker for early detection of adhd using phase-based brain connectivity and graph theory, Physical and Engineering Sciences in Medicine (2023) 1--19.

\bibitem{lin2023measurement}
G.~Lin, A.~Lin, Y.~Mi, D.~Gu, Measurement of information transfer based on phase increment transfer entropy, Chaos, Solitons \& Fractals 174 (2023) 113864.

\bibitem{gu2023detection}
D.~Gu, A.~Lin, G.~Lin, Detection of attention deficit hyperactivity disorder in children using ceemdan-based cross frequency symbolic convergent cross mapping, Expert Systems with Applications 226 (2023) 120105.

\bibitem{bakhtyari2022adhd}
M.~Bakhtyari, S.~Mirzaei, Adhd detection using dynamic connectivity patterns of eeg data and convlstm with attention framework, Biomedical Signal Processing and Control 76 (2022) 103708.

\bibitem{sanchis2024novel}
J.~Sanchis, S.~Garc{\'\i}a-Ponsoda, M.~A. Teruel, J.~Trujillo, I.-Y. Song, A novel approach to identify the brain regions that best classify adhd by means of eeg and deep learning, Heliyon 10~(4) (2024).

\bibitem{atila2023lsgp}
O.~Atila, E.~Deniz, A.~Ari, A.~Sengur, S.~Chakraborty, P.~D. Barua, U.~R. Acharya, Lsgp-usfnet: automated attention deficit hyperactivity disorder detection using locations of sophie germain’s primes on ulam’s spiral-based features with electroencephalogram signals, Sensors 23~(16) (2023) 7032.

\bibitem{abbas2021effective}
A.~K. Abbas, G.~Azemi, S.~Amiri, S.~Ravanshadi, A.~Omidvarnia, Effective connectivity in brain networks estimated using eeg signals is altered in children with adhd, Computers in Biology and Medicine 134 (2021) 104515.

\bibitem{khare2022vhers}
S.~K. Khare, N.~B. Gaikwad, V.~Bajaj, Vhers: a novel variational mode decomposition and hilbert transform-based eeg rhythm separation for automatic adhd detection, IEEE Transactions on Instrumentation and Measurement 71 (2022) 1--10.

\bibitem{khare2023explainable}
S.~K. Khare, U.~R. Acharya, An explainable and interpretable model for attention deficit hyperactivity disorder in children using eeg signals, Computers in biology and medicine 155 (2023) 106676.

\bibitem{maniruzzaman2023optimal}
M.~Maniruzzaman, M.~A.~M. Hasan, N.~Asai, J.~Shin, Optimal channels and features selection based adhd detection from eeg signal using statistical and machine learning techniques, IEEE Access 11 (2023) 33570--33583.

\bibitem{loh2023adhd}
H.~W. Loh, C.~P. Ooi, S.~L. Oh, P.~D. Barua, Y.~R. Tan, U.~R. Acharya, D.~S.~S. Fung, Adhd/cd-net: automated eeg-based characterization of adhd and cd using explainable deep neural network technique, Cognitive Neurodynamics (2023) 1--17.

\bibitem{kasim2023identification}
{\"O}.~Kasim, Identification of attention deficit hyperactivity disorder with deep learning model, Physical and Engineering Sciences in Medicine (2023) 1--10.

\bibitem{ge2023symbolic}
X.~Ge, A.~Lin, Symbolic convergent cross mapping based on permutation mutual information, Chaos, Solitons \& Fractals 167 (2023) 112992.

\bibitem{barua2022tmp19}
P.~D. Barua, S.~Dogan, M.~Baygin, T.~Tuncer, E.~E. Palmer, E.~J. Ciaccio, U.~R. Acharya, Tmp19: A novel ternary motif pattern-based adhd detection model using eeg signals, Diagnostics 12~(10) (2022) 2544.

\bibitem{ekhlasi2021direction}
A.~Ekhlasi, A.~M. Nasrabadi, M.~R. Mohammadi, Direction of information flow between brain regions in adhd and healthy children based on eeg by using directed phase transfer entropy, Cognitive Neurodynamics 15~(6) (2021) 975--986.

\bibitem{talebi2022investigating}
N.~Talebi, A.~M. Nasrabadi, Investigating the discrimination of linear and nonlinear effective connectivity patterns of eeg signals in children with attention-deficit/hyperactivity disorder and typically developing children, Computers in Biology and Medicine 148 (2022) 105791.

\bibitem{chauhan2023regional}
N.~Chauhan, B.-J. Choi, Regional contribution in electrophysiological-based classifications of attention deficit hyperactive disorder (adhd) using machine learning, Computation 11~(9) (2023) 180.

\bibitem{acharya2016american}
J.~N. Acharya, A.~J. Hani, J.~Cheek, P.~Thirumala, T.~N. Tsuchida, American clinical neurophysiology society guideline 2: guidelines for standard electrode position nomenclature, The Neurodiagnostic Journal 56~(4) (2016) 245--252.

\bibitem{kothe2014artifact}
C.~A.~E. Kothe, T.-P. Jung, Artifact removal techniques with signal reconstruction. 2016, Google Patents (2014).

\bibitem{chang2018evaluation}
C.-Y. Chang, S.-H. Hsu, L.~Pion-Tonachini, T.-P. Jung, Evaluation of artifact subspace reconstruction for automatic eeg artifact removal, in: 2018 40th Annual International Conference of the IEEE Engineering in Medicine and Biology Society (EMBC), IEEE, 2018, pp. 1242--1245.

\bibitem{mullen2015real}
T.~R. Mullen, C.~A. Kothe, Y.~M. Chi, A.~Ojeda, T.~Kerth, S.~Makeig, T.-P. Jung, G.~Cauwenberghs, Real-time neuroimaging and cognitive monitoring using wearable dry eeg, IEEE transactions on biomedical engineering 62~(11) (2015) 2553--2567.

\bibitem{delorme2004eeglab}
A.~Delorme, S.~Makeig, Eeglab: an open source toolbox for analysis of single-trial eeg dynamics including independent component analysis, Journal of neuroscience methods 134~(1) (2004) 9--21.

\bibitem{makoto2021cleanrawdata}
A.~D. Makoto~Miyakoshi, S.~Makeig, Cleanrawdata eeglab plugin, \url{https://eeglab.org/others/EEGLAB_Extensions.html}, [Online; last accessed 2022-02-04] (2021).

\bibitem{makeig1995independent}
S.~Makeig, A.~Bell, T.-P. Jung, T.~J. Sejnowski, Independent component analysis of electroencephalographic data, Advances in neural information processing systems 8 (1995).

\bibitem{pion2019iclabel}
L.~Pion-Tonachini, K.~Kreutz-Delgado, S.~Makeig, Iclabel: An automated electroencephalographic independent component classifier, dataset, and website, NeuroImage 198 (2019) 181--197.

\bibitem{klug2024optimizing}
M.~Klug, T.~Berg, K.~Gramann, Optimizing eeg ica decomposition with data cleaning in stationary and mobile experiments, Scientific Reports 14~(1) (2024) 14119.

\bibitem{reddy2024nonlinear}
C.~S. Reddy, M.~R. Reddy, Nonlinear difference subspace method of motor imagery eeg classification in brain-computer interface, Digital Signal Processing 155 (2024) 104720.

\bibitem{zhang2024adaptive}
X.~Zhang, Y.~Wang, Y.~Tang, Z.~Wang, Adaptive filter of frequency bands based coordinate attention network for eeg-based motor imagery classification, Health Information Science and Systems 12~(1) (2024) 11.

\bibitem{schiratti2018ensemble}
J.-B. Schiratti, J.-E. Le~Douget, M.~Le~Van~Quyen, S.~Essid, A.~Gramfort, An ensemble learning approach to detect epileptic seizures from long intracranial eeg recordings, in: 2018 IEEE International Conference on Acoustics, Speech and Signal Processing (ICASSP), IEEE, 2018, pp. 856--860.

\bibitem{paivinen2005epileptic}
N.~P{\"a}ivinen, S.~Lammi, A.~Pitk{\"a}nen, J.~Nissinen, M.~Penttonen, T.~Gr{\"o}nfors, Epileptic seizure detection: A nonlinear viewpoint, Computer methods and programs in biomedicine 79~(2) (2005) 151--159.

\bibitem{esteller2001comparison}
R.~Esteller, G.~Vachtsevanos, J.~Echauz, B.~Litt, A comparison of waveform fractal dimension algorithms, IEEE Transactions on Circuits and Systems I: Fundamental Theory and Applications 48~(2) (2001) 177--183.

\bibitem{esteller2001line}
R.~Esteller, J.~Echauz, T.~Tcheng, B.~Litt, B.~Pless, Line length: an efficient feature for seizure onset detection, in: 2001 Conference Proceedings of the 23rd Annual International Conference of the IEEE Engineering in Medicine and Biology Society, Vol.~2, IEEE, 2001, pp. 1707--1710.

\bibitem{demanuele2007distinguishing}
C.~Demanuele, C.~J. James, E.~J. Sonuga-Barke, Distinguishing low frequency oscillations within the 1/f spectral behaviour of electromagnetic brain signals, Behavioral and Brain Functions 3 (2007) 1--14.

\bibitem{winkler2011automatic}
I.~Winkler, S.~Haufe, M.~Tangermann, Automatic classification of artifactual ica-components for artifact removal in eeg signals, Behavioral and brain functions 7 (2011) 1--15.

\bibitem{qian2004hurst}
B.~Qian, K.~Rasheed, Hurst exponent and financial market predictability, in: IASTED conference on Financial Engineering and Applications, Proceedings of the IASTED International Conference Cambridge, MA, 2004, pp. 203--209.

\bibitem{devarajan2014eeg}
K.~Devarajan, E.~Jyostna, K.~Jayasri, V.~Balasampath, Eeg-based epilepsy detection and prediction, International Journal of Engineering and Technology 6~(3) (2014) 212.

\bibitem{richman2000physiological}
J.~S. Richman, J.~R. Moorman, Physiological time-series analysis using approximate entropy and sample entropy, American journal of physiology-heart and circulatory physiology 278~(6) (2000) H2039--H2049.

\bibitem{inouye1991quantification}
T.~Inouye, K.~Shinosaki, H.~Sakamoto, S.~Toi, S.~Ukai, A.~Iyama, Y.~Katsuda, M.~Hirano, Quantification of eeg irregularity by use of the entropy of the power spectrum, Electroencephalography and clinical neurophysiology 79~(3) (1991) 204--210.

\bibitem{teixeira2011epilab}
C.~Teixeira, B.~Direito, H.~Feldwisch-Drentrup, M.~Valderrama, R.~Costa, C.~Alvarado-Rojas, S.~Nikolopoulos, M.~Le~Van~Quyen, J.~Timmer, B.~Schelter, et~al., Epilab: A software package for studies on the prediction of epileptic seizures, Journal of Neuroscience Methods 200~(2) (2011) 257--271.

\bibitem{kharbouch2011algorithm}
A.~Kharbouch, A.~Shoeb, J.~Guttag, S.~S. Cash, An algorithm for seizure onset detection using intracranial eeg, Epilepsy \& Behavior 22 (2011) S29--S35.

\bibitem{mormann2007seizure}
F.~Mormann, R.~G. Andrzejak, C.~E. Elger, K.~Lehnertz, Seizure prediction: the long and winding road, Brain 130~(2) (2007) 314--333.

\bibitem{badani2017detection}
S.~Badani, S.~Saha, A.~Kumar, S.~Chatterjee, R.~Bose, Detection of epilepsy based on discrete wavelet transform and teager-kaiser energy operator, in: 2017 IEEE Calcutta Conference (CALCON), IEEE, 2017, pp. 164--167.

\bibitem{degirmenci2023statistically}
M.~Degirmenci, Y.~K. Yuce, M.~Perc, Y.~Isler, Statistically significant features improve binary and multiple motor imagery task predictions from eegs, Frontiers in Human Neuroscience 17 (2023) 1223307.

\bibitem{d1971omnibus}
R.~B. d'Agostino, An omnibus test of normality for moderate and large size samples, Biometrika 58~(2) (1971) 341--348.

\bibitem{d1973tests}
R.~D'agostino, E.~S. Pearson, Tests for departure from normality. empirical results for the distributions of b 2 and sqrt(b), Biometrika 60~(3) (1973) 613--622.

\bibitem{chen2016xgboost}
T.~Chen, C.~Guestrin, Xgboost: A scalable tree boosting system, in: Proceedings of the 22nd acm sigkdd international conference on knowledge discovery and data mining, 2016, pp. 785--794.

\bibitem{deshmukh2024contributions}
M.~Deshmukh, M.~Khemchandani, P.~M. Thakur, Contributions of brain regions to machine learning-based classifications of attention deficit hyperactivity disorder (adhd) utilizing eeg signals, Applied Neuropsychology: Adult (2024) 1--15.

\bibitem{karimui2022adhd}
R.~Y. Karimui, G.~S. Bajestani, B.~Sheikholeslami, The adhd effects on partial opposites in trigonometric plots obtained from the eeg signals, Chaos, Solitons \& Fractals 158 (2022) 112021.

\bibitem{sharma2023attention}
Y.~Sharma, B.~K. Singh, Attention deficit hyperactivity disorder detection in children using multivariate empirical eeg decomposition approaches: A comprehensive analytical study, Expert Systems with Applications 213 (2023) 119219.

\bibitem{vallat2021open}
R.~Vallat, M.~P. Walker, An open-source, high-performance tool for automated sleep staging, Elife 10 (2021) e70092.

\bibitem{taghibeyglou2022detection}
B.~TaghiBeyglou, A.~Shahbazi, F.~Bagheri, S.~Akbarian, M.~Jahed, Detection of adhd cases using cnn and classical classifiers of raw eeg, Computer Methods and Programs in Biomedicine Update 2 (2022) 100080.

\end{thebibliography}



\end{document}